\documentclass[12pt,a4paper]{article}

\usepackage{xcolor}
\usepackage{bm}
\usepackage{graphicx}
\usepackage{amsmath}
\usepackage{amssymb}
\usepackage{amsbsy}
\usepackage{amsthm}
\usepackage{amsfonts}
\usepackage{bbm}
\usepackage{amsfonts}
\usepackage{slashed}
\usepackage{subfigure}
\usepackage{jheppub}

\def\bea{\begin{eqnarray}}
\def\eea{\end{eqnarray}}
\def\nn{\nonumber}
\def\ba{\begin{array}}
	\def\ea{\end{array}}
\def\nn{\nonumber}
\def\Tr{\text{Tr}}
\def\sgn{\text{sgn}}
\def\J{\mathcal{J}}

\def\L{\mathcal{L}}

\def\I{\mathbbm{I}}
\def\H{\mathcal{H}}

\def\K{\mathcal{K}}

\def\avg#1{\left\langle#1\right\rangle}
\def\bra#1{\left\langle#1\right|}
\def\ket#1{\left|#1\right\rangle}

\def\abs#1{\left|#1\right|}
\def\kc#1{\left(#1\right)}
\def\kd#1{\left[#1\right]}
\def\ke#1{\left\{#1\right\}}
\def\Re{{\rm Re}}
\def\Im{{\rm Im}}
\def\sgn{{\rm sgn}}

\def\be{\begin{equation}}       \def\ee{\end{equation}}
\def\bea{\begin{eqnarray}}      \def\eea{\end{eqnarray}}
\def\ba{\begin{array}}
	\def\ea{\end{array}}
\def\bnum{\begin{enumerate} }
	\def\enum{\end{enumerate}}

\def\nn{\nonumber}

\def\=>{\Rightarrow}
\def\>{\rightarrow}

\def\eye2{Fathbb{I}}

\def\Tr{\mathrm{Tr}}

\DeclareMathOperator\sinc{sinc}

\DeclareMathOperator\Si{Si}

\title{\Large Universal chaotic dynamics from Krylov space}

\author[a]{Johanna~Erdmenger,}
\author[b]{Shao-Kai~Jian,}
\author[a,1]{and Zhuo-Yu~Xian\note{Corresponding author.}
\note{Authors' names are listed in  alphabetical order.}}

\emailAdd{erdmenger@physik.uni-wuerzburg.de}
\emailAdd{sjian@tulane.edu} 
\emailAdd{zhuo-yu.xian@physik.uni-wuerzburg.de}

\affiliation[a]{Institute for Theoretical Physics and Astrophysics and W{\"u}rzburg-Dresden Cluster of \\ Excellence ct.qmat,
	Julius-Maximilians-Universit{\"a}t W{\"u}rzburg,\\ D-97074 W{\"u}rzburg, Germany}
\affiliation[b]{Department of Physics and Engineering Physics, Tulane University, \\New Orleans, Louisiana, 70118, USA}

\abstract{
Krylov complexity measures the spread of the wavefunction in the Krylov basis, which is constructed using the Hamiltonian and an initial state. We investigate the evolution of the maximally entangled state in the Krylov basis for both chaotic and non-chaotic systems. For this purpose, we derive an Ehrenfest theorem for the Krylov complexity, which reveals its close relation to the spectrum. Our findings suggest that neither the linear growth nor the saturation of Krylov complexity is necessarily associated with chaos. However, for chaotic systems, we observe a universal rise-slope-ramp-plateau behavior in the transition probability from the initial state to one of the Krylov basis states. 
Moreover, a long ramp in the transition probability is a signal for spectral rigidity, characterizing quantum chaos. Also, this ramp is directly responsible for the late-time peak of Krylov complexity observed in the literature. On the other hand, for non-chaotic systems, this long ramp is absent.
Therefore, our results help to clarify which features of the wave function time evolution in Krylov space characterize chaos. We exemplify this by considering the Sachdev-Ye-Kitaev model with two-body or four-body interactions.
}

\begin{document}
\maketitle
\flushbottom

\section{Introduction}

\subsection{Outline and results}

\noindent Krylov complexity measures the spread of a time-evolving state in a Hilbert space. 
For a maximally entangled state, this complexity only depends on the spectrum of the Hamiltonian and is independent of the choice of fundamental gates. 
We study the universal behavior of Krylov complexity for Hamiltonians describing chaotic systems.
The Krylov approach consists of defining a particular Hilbert space basis, the Krylov basis. 
State evolution in this basis can be mapped to a particle moving on a one-dimensional chain. 
We exploit this map to equivalently describe state evolution in the Krylov basis in terms of forces acting on the particle.
At early times, Krylov complexity displays a linear growth.
Here we find that this linear growth is described by a generalized  Ehrenfest theorem in Krylov space, providing an effective classical equation of motion for Krylov complexity.
The linear growth of Krylov state complexity is not a characteristic of chaos, since it may also appear in non-chaotic systems. 
For late times, in chaotic systems Krylov complexity shows a characteristic peak and saturation structure, as numerically observed in Ref.~\cite{Balasubramanian:2022tpr}.
By taking the continuum limit of the one-dimensional chain, we derive an analytical expression for the Krylov complexity at late times that confirms the observation of Ref.~\cite{Balasubramanian:2022tpr}.
Moreover, we calculate the wave function in the Krylov basis. 
The norm of the amplitude squared of this wave function is referred to as transition probability. 
We find that it exhibits a universal rise-slope-ramp-plateau behavior with a long ramp. 
The ramp-plateau behavior is characteristic for chaos. Due to probability conservation,
the characteristic long ramp that we find gives rise to the peak structure of the Krylov complexity. 
Moreover, we find that in  non-chaotic systems, the long ramp of the transition probability disappears. 
This implies that in this case, the peak in the Krylov complexity is absent. Our results thus clarify which features of the wave function time evolution in Krylov space characterize chaos.   

\subsection{Chaos and Krylov space}

\noindent To put our results into context, we begin with a brief review of quantum chaotic systems and the Krylov approach. 

The time evolution of a quantum chaotic system is characterized by the statistics of the energy spectrum. 
In a quantum chaotic system, the energy levels are correlated and exhibit two salient phenomena: level repulsion and spectral rigidity \cite{Bohigas:1983er,Berry:1985semiclassical,Muller:2004semiclassical}. 
Level repulsion refers to the fact that energy levels tend to avoid clustering.
Spectral rigidity means that the number of levels within an energy interval of given size has small fluctuations. 
Both of these properties are due to the precise nature of correlations between level spacings in chaotic systems.
More precisely, it is expected that the level spacing statistics coincides with the random matrix theories (RMT) and are well-approximated by the Wigner-Dyson distribution \cite{dyson1962statisticalI,dyson1962statisticalII,dyson1972class,Guhr:1997ve}. 
The most studied RMT is the $\tilde\beta$-Gaussian ensemble \cite{Dumitriu:2002beta}, which we will focus on in this paper. 
Recently, the Sachdev-Ye-Kitaev (SYK) model \cite{Kitaev:2015a,Maldacena:2016remarks}, a quantum mechanical model with all-to-all interactions (as opposed to nearest-neighbor interactions), was found to exhibit the level statistics of Gaussian RMT \cite{you2017sachdev,Garcia-Garcia:2016mno}. 
Moreover, Jackiw-Teitelboim (JT) gravity, a two-dimensional dilaton gravity that shows similar features of the SYK model, is precisely consistent with a double-scaled random matrix integral \cite{Saad:2019lba}.

\begin{figure}
    \centering
    \includegraphics[height=0.3\linewidth]{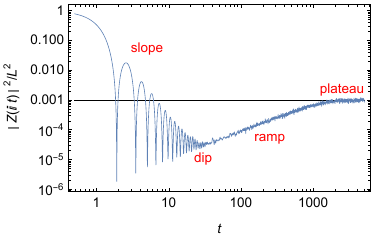}
    ~~\includegraphics[height=0.3\linewidth]{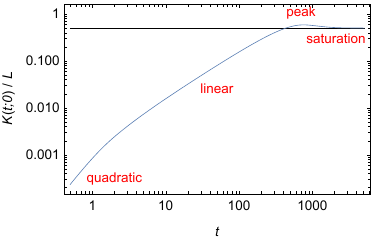}
    \caption{The SFF (left) and Krylov complexity of maximally entangled state (right) as functions of time for the Gaussian unitary ensemble with dimension $L=1024$ and $128$ realizations. The blue curves represent the numerical results and the black lines denote the values of $1/L$ (left) and $1/2$ (right).}
    \label{fig:SFFKCGUETFD}
\end{figure}

The level statistics also determines the behavior of the spectral form factor (SFF) given by the square of the absolute value of the partition function with a complex time argument \cite{brezin1997spectral,prange1997spectral},
\begin{align}\label{SFF}
    \abs{Z(\beta+it)}^2=\abs{\Tr\,e^{-(\beta+it)H}}^2=\sum_{p,q=0}^{D-1} e^{-\beta(E_p+E_q)-it(E_p-E_q)} \, ,
\end{align}
where $E_p$ and $D$ are the $p$-th eigenvalue and the dimension of the Hamiltonian $H$, respectively, and $\beta,\, t \in \mathbb R$.
In a chaotic system, the SFF for the ensemble average usually shows three regions as a function of time: slope, ramp, and plateau \cite{Cotler:2016fpe,Cotler:2017jue,Liu:2018hlr}, as shown in Fig.~\ref{fig:SFFKCGUETFD}. 
Roughly speaking, these features arise from the width of the spectrum, spectral rigidity, and level repulsion, respectively. 
Since its time evolution reflects these properties, the SFF may be used to diagnose quantum chaos. 
The time at which there is a cross-over between the slope and ramp evolution is referred to as dip time.
A chaotic system usually has an exponentially late dip time and an exponentially long and linear ramp region, controlled by the long-range spectral rigidity~\cite{dyson1962statisticalI,Guhr:1997ve,Cotler:2016fpe}. 

Chaotic evolution is a complicated process that requires a complexity measure for its quantitative analysis.   
Partially motivated by new relations between quantum computation and the time evolution of black holes \cite{lloyd2000ultimate,Susskind:2014rva,Harlow:2022qsq,Faulkner:2022mlp}, several concepts of complexity were proposed to measure how many computational steps are required to reach a target state or operator from a reference state or operator. 
One of the motivations for the investigations in the present paper is to examine how complexity reflects late-time chaos, based on level repulsion and spectral rigidity. 

Among the complexity measures in information theory, Nielsen defined the complexity of a unitary operator $U(t)=e^{-iHt}$ as the minimal distance to the identity in the unitary group \cite{nielsen2005geometric,nielsen2006quantum,dowling2006geometry}. 
The minimal distance on the group manifold is defined in terms of some cost function. 
The definition of the cost relies on the choice of few- or many-body terms based on the locality properties of the Hamiltonian $H$.
A similar choice of fundamental operations appears in the notion of computational complexity, which measures the complexity of producing a target state $\ket{\psi_T}$ starting from a reference state $\ket{\psi_R}$ \cite{Aaronson:2016vto,watrous2008quantum}. 
Given a set of elementary quantum gates, the computational complexity is the minimum number of elementary gates necessary to achieve a unitary transformation $U$ within a precision so that $\ket{\psi_T}=U\ket{\psi_R}$. 
Here, the ambiguity in defining complexity is related to the choice of gates. 
Both Nielsen's complexity and computational complexity were investigated for free many-body systems and field theories \cite{Jefferson:2017sdb,Chapman:2017rqy,Hackl:2018ptj,Molina-Vilaplana:2018sfn,Khan:2018rzm}, interacting systems  \cite{Balasubramanian:2019quantum,Ali:2019zcj,Bhattacharyya:2020iic,Bhattacharyya:2019txx,Magan:2018nmu,Caputa:2018kdj,Erdmenger:2020sup}, and for large qudit systems \cite{Basteiro:2021ene,Lv:2023jbv}. 

When discussing the relation between late-time chaos and these notions of complexity, we note that level statistics does not provide the information about the locality properties of the Hamiltonian directly. 
Given a Hamiltonian from a Gaussian matrix ensemble, there is no natural way to define locality \cite{Roberts:2016design,Cotler:2017jue}, let alone few-body or many-body interaction terms. 
As we will describe below, Krylov complexity is unambiguously defined even in this case, and hence well-suited for matrix ensembles.

Notions of complexity were also proposed in the context of the AdS/CFT correspondence \cite{Maldacena:1997re}. 
In particular, the volume or action of a wormhole connecting the two sides of an eternal black hole is conjectured to be related to the complexity of preparing the dual state in quantum field theory
\cite{Susskind:2014switchback,Stanford:2014complexity,Susskind:2014rva,Brown:2015action,Susskind:2019newton,Susskind:2020momentum}. 
The real-time evolution of holographic complexity exhibits a similar linear-to-plateau behavior as the computational complexity \cite{Brown:2017secondlaw}, where the growth rate is argued to be bounded by the energy \cite{lloyd2000ultimate}. 
Moreover, holographic complexity has a nonzero initial value that is proportional to the initial volume of the wormhole in the dual gravity theory. 
The eternal black hole corresponds to a thermofield double (TFD) state in the field theory \cite{Maldacena:2001eternal}. 
So, the initial volume as well as the complexity are generated by imaginary time evolution in preparing the TFD state. 
So far, a precise holographic dual of complexity is still an open question, despite recent progress \cite{Abt:2017pmf,Chapman:2018hou,Brown:2018falling,Susskind:2018fall,Lin:2019schwarzian,Susskind:2019newton,Susskind:2020momentum,Brown:2018JT}. 
One of the remaining challenges is to precisely define complexity for interacting quantum field theories, in particular since their Hilbert space is infinite dimensional. 
One approach in this direction is to consider CFTs and to construct gates from conformal symmetry transformations \cite{Caputa:2018kdj,Erdmenger:2020sup,Flory:2020eot,Flory:2020dja, Chagnet:2021uvi, Erdmenger:2021wzc}. This also allows to construct a gravity dual of the cost function \cite{Erdmenger:2022lov}.

With potential relations to holography in mind,  the notion of Krylov complexity draws increasing attention \cite{Parker:2018a,Barbon:2019on,Rabinovici:2020operator,Jian:2020qpp,Dymarsky:2021bjq,Caputa:2021sib,Balasubramanian:2022tpr,Balasubramanian:2022dnj} since it is well-defined in any quantum theory.
Krylov complexity has the advantage that its complexity measure is independent of the locality properties of the Hamiltonian. 
It does not rely on defining elementary gates or a given tolerance. 
This makes it very appealing in the context of holographic dualities.
According to the Hilbert space on which Krylov complexity is defined, it describes the evolution of states \cite{Balasubramanian:2022tpr} or  operators \cite{Parker:2018a}, in both real time and imaginary time \cite{Dymarsky:2019quantum,Avdoshkin:2019trj}.

In \cite{Balasubramanian:2022tpr}, a notion of Krylov {\it state} complexity is defined that realizes the appealing visualization of a wavefunction spreading over the Hilbert space in a basis-independent way. 
The authors of~\cite{Balasubramanian:2022tpr} refer to this Krylov state complexity as `spread complexity'.
It measures how far the target state spreads in the Hilbert space $\H$. 
The target state $\ket{\psi_\tau}=e^{-\tau\L}\ket0$ starts from a reference state $\ket{0}$ at $\tau =0$ and evolves under a Liouvillian operator $\L$ constructed from the Hamiltonian $H$.
Based on the Taylor series of $\L$, this evolution may be studied in {\it Krylov space} that is constructed by applying $\L$ on $\ket0$ repeatedly. 
In the orthogonal and normalized basis of Krylov space, namely $\ke{\ket{O_n}}$ with $\ket{O_n}=\psi_n(\L)\ket0$ and $\psi_n(x)$ a polynomial of degree $n$, the Liouvillian $\L$ becomes a tridiagonal matrix, whose components are called Lanczos coefficients \cite{viswanath2008recursion,lanczos1950iteration}, denoted as $\ke{a_n,b_n}$. 
In terms of the Krylov basis, the time evolution of a state $\ket{\psi_\tau}=e^{-\tau\L}\ket0$ can be effectively mapped to the propagation of a quantum particle along a one-dimensional chain, which is referred to as Krylov chain \cite{Parker:2018a}. 
Krylov complexity is then defined as the location of the particle in the Krylov chain. 
This is equivalent to the  expected number of times of applying $\L$ on $\ket0$ required to generate $\ket{\psi_\tau}$.

Krylov {\it operator} complexity measures how far an operator in the Heisenberg picture spreads in the space of operators. 
By the Gelfand–Naimark–Segal (GNS) construction \cite{gelfand1994imbedding,segal1947irreducible,Magan:2020iac}, the space of operators is isometric to a double-copy Hilbert space. 
More precisely, the reference state is defined as $\ket{O}=(O\otimes\I)\ket0$, with $O$ an arbitrary operator and $\I$ the identity, acting on the single-copy Hilbert space, respectively, and $\ket0$ is a maximally entangled state in the double-copy Hilbert space. 
Moreover, one considers a Liouvillian $\L=H\otimes\I-\I\otimes H$, where $H$ is the Hamiltonian acting on the single-copy Hilbert space. 
Since $\L\ket0=0$, the application of $\L$ on $\ket O$ is nothing but the commutator, namely $\L\ket O= ([H,O]\otimes\I)\ket0$. 
Then, the Krylov operator complexity $e^{-iHt}Oe^{iHt}$ is identified as the Krylov state complexity $e^{-it\L}\ket O$ in the double-copy Hilbert space. 
Once the operator $O$ has a nonzero commutator with the Hamiltonian, it will grow under the evolution with $\L$. 
Methods for studying the time evolution of Krylov complexity were recently obtained by decomposing Liouvillian $\L$ into annihilation and creation operators and analyzing the ``complexity algebra'' \cite{Caputa:2021sib,Haque:2022ncl,Bhattacharjee:2022qjw,Hornedal:2022pkc}.

The exponential growth of Krylov operator complexity, and also the linear growth of Lanczos coefficients, allow to obtain the Lyapunov exponent~\cite{Parker:2018a,Barbon:2019wsy} which characterizes the exponential operator size growth  \cite{Roberts:2018operator,Qi:2018quantum} given by, {\it e.g.}, the out-of-time-ordered correlator (OTOC) \cite{Shenker:2013black,Roberts:2014localized,Mertens:2017solving,Shenker:2013yza}. 
However, the maximally exponential growth of Krylov operator complexity at early times is also observed in integrable systems, including free field theories. 
Exponential growth is therefore not necessarily related to chaos \cite{Dymarsky:2021bjq,Bhattacharjee:2022vlt}.
Here, we hence also turn our attention to the relation between the late-time behavior of Krylov complexity and chaos \cite{Rabinovici:2020operator,Espanol:2022cqr}. 
Moreover, it is argued in \cite{Kar:2021nbm,Rabinovici:2021qqt,Rabinovici:2022beu,Alishahiha:2022anw} that the descent in the Lanczos coefficients as well as the late-time behavior of the Krylov operator complexity given by the evolution with a chaotic Hamiltonian $H$ is expected to be governed by the RMT.
The relation between Krylov complexity and chaos has further been discussed for a number of models, including the SYK models \cite{Jian:2020qpp,Bhattacharjee:2022ave,He:2022ryk}, quantum field theories \cite{Caputa:2021ori,Khetrapal:2022dzy,Kundu:2023hbk,Camargo:2022rnt,Avdoshkin:2022xuw}, many-body localization system \cite{Trigueros:2021rwj}, and open systems \cite{Liu:2022god,Bhattacharya:2022gbz,Bhattacharjee:2022lzy,Bhattacharya:2023zqt}. 
Krylov complexity has also been used for distinguishing topological phases \cite{Caputa:2022yju,Caputa:2022eye} and for investigating the quantum charging advantage of SYK-like quantum batteries \cite{Kim:2021okd}.

To study late-time chaos from Krylov complexity, it appears to be more convenient to study Krylov state complexity directly.
The Krylov complexity of the maximally entangled state only depends on the spectrum of the Hamiltonian $H$. 
It is thus tied to the SFF and suitable to describe late-time chaos \cite{Balasubramanian:2022tpr}, in particular during the time range when the chaotic level spacing becomes manifest \cite{Rabinovici:2020operator,Kar:2021nbm}. 
In Fig.~\ref{fig:SFFKCGUETFD}, we show the Krylov state complexity of the maximally entangled state, and its correspondence to the SFF. 
It exhibits quadratic growth, linear growth, a peak, and saturation, whose transition time scales are close to those in the SFF. We refer to the time when it reaches its peak as ``peak time''.
We further refer to the quadratic growth and linear growth regions as the early-time behavior and to the peak and saturation as the late-time behavior. 
We will show that the early-time behavior is given by a double time integral of the SFF via an Ehrenfest theorem and the late-time behavior is determined by the universal behavior of the probability given by the wave function in the Krylov space.

\subsection{Organization of the paper}

\noindent
In Sec.~\ref{sec:Krylov}, we first review the construction of Krylov space and Krylov complexity. 
We introduce the continuum limit of the Krylov approach in a first-order and a second-order formalism, respectively. 
Moreover, we determine the Krylov complexity for obtaining the TFD state from a reference state given by a maximally entangled state. 
This Krylov complexity is entirely determined by the Hamiltonian spectrum.

In Sec.~\ref{sec:Earlytime}, we consider the Gaussian matrix ensemble and study the evolution of Krylov complexity at early times. 
We propose an Ehrenfest theorem for Krylov complexity, which linearly relates the second-order time derivative of the Krylov complexity to the SFF, see \eqref{SFF}. 
In particular, with the Lanczos coefficients given by Gaussian matrix ensemble, the linear growth of the Krylov complexity is the determined by the slope of SFF, which is not necessarily related to chaos.

In Sec.~\ref{sec:chaos}, we study the evolution of a maximally entangled state in Krylov space at late times. 
For the Gaussian unitary ensemble (GUE), we numerically study the distribution and evolution of the {\it transition probability} $|\avg{O_n|e^{-\tau\L}|0}|^2$, namely the probability for reaching each state in the Krylov basis. 
We find that the transition probability universally exhibits a rise-slope-ramp-plateau behavior with an exponentially long ramp. 
Like the SFF, the ramp-plateau behavior exhibited in the transition probability characterizes chaos. 
To analytically explain and estimate this behavior, we further approximate the polynomial $\psi_n(\L)$ and derive an expression for the rise-slope-ramp-plateau behavior in App.~\ref{sec:PolynomialEnsemble}. 
Moreover, we show that the above ramp-plateau behavior generally appears in any subspace observable in the Krylov space. 
Finally, we show that the existence of the long ramp in the transition probability is directly responsible for the peak in the Krylov complexity.

In Sec.~\ref{sec:nonChaotic}, we study the transition probability and Krylov complexity for a non-chaotic spectrum, where the levels are uncorrelated. 
In contrast to the chaotic case, the transition probability here exhibits a rise-slope-plateau behavior without a ramp. 
The absence of a ramp in the transition probability is directly responsible for the absence of a peak in Krylov complexity. 
However, the linear growth of Krylov complexity persists.

In Sec.~\ref{sec:SYK}, we further study the transition probability and Krylov complexity in the SYK model. 
In the SYK$_4$ model, the transition probability exhibits a rise-slope-ramp-plateau behavior similar to the case of RMT. 
In the SYK$_2$ model, the  transition probability exhibits a rise-slope-ramp-plateau behavior with a short ramp at small $n$ and no ramp at large $n$.

We conclude in  Sec.~\ref{sec:conclusion} with an outlook to future directions.

\section{Krylov state complexity}\label{sec:Krylov}

In this section, we detail the general framework for Krylov space and Krylov complexity of a time-evolving state.

\subsection{Krylov space}

Given a Hilbert space $\H$, a reference state $\ket{0}\in \H$, and a Hermitian operator $\L$ acting on $\H$ called Liouvillian \cite{Parker:2018a}, we can construct the Krylov space as follows. 
First, we construct a sequence of normalized states $\ke{\ket{A_j}}_{j=0}^{K-1}$ by subsequently applying $\L$ to $\ket0$, namely
$ 
\ket{A_j}	=\mu_{2j}^{-1/2}\L^j\ket{0},	
$ 
for $ j=0,1,2,\cdots,K-1$, with
$
	\mu_{j}=\bra0 \L^{j}\ket0
$
the moments and $K$ the minimal number such that $\mu_{2K}=0$. 
In general, these states may not be independent of each other. 
Let $L$ to be the index of the first state $\ket{A_L}$ becoming linearly dependent on the former states $\ke{\ket{A_j}}_{j=0}^{L-1}$. 
Then its latter states $\ke{\ket{A_j}}_{j=L}^{K-1}$ are also linearly dependent on these former states $\ke{\ket{A_j}}_{j=0}^{L-1}$. 
So $\ke{\ket{A_j}}_{j=0}^{K}$ only span a $L$-dimensional space $\K$, called Krylov space. 
Usually, we take the set of the former $L$ states $\ke{\ket{A_j}}_{j=0}^{L-1}$ as its basis.

In general, the states $\ke{\ket{A_j}}_{j=0}^{L-1}$ are not orthogonal to each other. 
We may apply the Gram-Schmidt orthogonalization to the sequence $\ke{\ket{A_j}}_{j=0}^{L-1}$ to generate a sequence of orthogonal states
\begin{align}\label{KrylovBasis}
	\ke{\ket{O_n}}_{n=0}^{L-1}, \quad 
	\ket{O_n}
	=\frac1{\sqrt{h_n}}p_n(\L)\ket{0}
	=\psi_n(\L)\ket{0},
\end{align}
where $p_n(x)$ and $\psi_n(x)$ are respectively the monic and normalized orthogonal polynomial of degree $n$ with a measure given by the spectrum of $\L$ \cite{Muck:2022xfc}. 
The norm $h_n$ will be determined later. 
Define the projection on the Krylov space as $\pi_\K=\sum_{n=0}^{L-1}\ket{O_n}\bra{O_n}$. 
Let $\ket{E_p}\in\K$ to be the eigenstate of $\L_\K=\pi_K\L\pi_K$, namely, $\pi_\K\L\ket{E_p}=E_p\ket{E_p}$ for $ p=0,1,\cdots,K-1$. The orthogonality relation and completeness relation are
\begin{align}\label{Orthogonal}
	&\avg{O_m|O_n}=\bra0 \psi_m(\L)\psi_n(\L)\ket0=\int_E \psi_m(E)\psi_n(E)=\delta_{mn}\\
	&\sum_n\avg{E_p|O_n}\avg{O_n|E_q}=\avg{E_p|0}\avg{0|E_q}\sum_n \psi_n(E_p)\psi_n(E_q)
	=\delta_{pq} \label{Complete}
\end{align}
where $\sum_n$ is the shorthand of $\sum_{n=0}^{L-1}$ and the measure in $\int_E$ on the spectrum $\ke{E_p}$ of $\L_\K$ is defined as
\begin{align}\label{Measure}
	\int_E f(E)
	\equiv\sum_p \abs{\avg{E_p|0}}^2 f(E_p)
	=\int dE \rho(E)\abs{\avg{E|0}}^2 f(E),
\end{align}
with the spectral density
\begin{align}\label{Density}
	\rho(\omega)=\sum_p \delta(\omega-E_p),
\end{align}
and a continuation of inner product $\abs{\avg{E_p|0}}^2$. 
Formally, we can also write the completeness relation as
$
	\sum_n \psi_n(E)\psi_n(E')=\bm\delta(E-E'),
$
where the function $\bm\delta(E-E')$ is defined as
$
	\int_E \bm \delta(E-E')f(E)=f(E').
$

The above Gram-Schmidt orthogonalization is realized by the following iterative algorithm \cite{viswanath2008recursion,lanczos1950iteration}
\begin{align}
	& \ket{O_0} = \ket{0}, \quad b_0=0,\nn\\
	&b_n\ket{O_n} = (\mathcal L-a_{n-1})\ket{O_{n-1}} - b_{n-1} \ket{O_{n-2}}, \quad 1\leq n \leq L-1,\label{Algorithm}\\
	& a_n=\bra{O_n}\mathcal L\ket{O_n},\quad \avg{O_n|O_n}=1.\nn
\end{align}
where $\{a_n,b_n\}$ are the Lanczos coefficients with the dimension of energy. 
By default, we choose $b_n\geq0$. 
The Gram-Schmidt orthogonalization is equivalent to a tridiagonalization of the Liouvillian $\L$ into a matrix $\mathbf L$,
\begin{align}
	L_{mn}=\bra{O_m}\L\ket{O_n}
	=\begin{pmatrix}
		a_0		&	b_1	&	0	&	\cdots	&	0	\\
		b_1		&	a_1	&	b_2	&	\cdots	&	0	\\
		0		&	b_2	&	a_2	&	\cdots	&	0	\\		\vdots	&\vdots	&\vdots	&   \ddots	&b_{L-1}\\
		0		&	0	&	0	&	b_{L-1}&a_{L-1}
	\end{pmatrix}.
\end{align}
The Lanczos coefficient can also be generated by the moments $\mu_j$ and vice versa.
The Lanczos coefficients give the monic polynomials,
\begin{align}\label{polynomial}
	p_n(E)=\det(E-\mathbf{L}^{(n)}),
\end{align}
where $\mathbf L^{(n)}$ is the $n\times n$ sub-matrix $\ke{L_{pq}}_{p,q=0}^{n-1}$. 
Obviously, the spectrum of $\L$ are the roots of $p_L(E)=0$. 
The norm $h_n$ of the monic polynomial $p_n(E)$ is given by
$
	b_n^2=h_n/h_{n-1}
$ and $h_0=1$.
The polynomials satisfy the recurrence relation. 
For the normalized polynomials,
\begin{align}\label{recurrence}
	E \psi_n(E)
	=\sum_m L_{nm}\psi_m(E)=b_{n+1}\psi_{n+1}(E)+a_n\psi_n(E)+b_n\psi_{n-1}(E),
\end{align}
where $b_0=b_L=0$. 
From \eqref{Orthogonal}, \eqref{Complete}, the $n$-th component of the eigenvector of $\mathbf L$ for energy $E_p$ is given by
$
v_n(E_p)=\avg{0|E_p}\psi_n(E_p)
$. 
Then the eigenstate of $\L_\K$ in $\K$ can be written as $\ket{E_p}=\sum_n v_n(E_p)\ket{O_n}$.

Given a (non-normalized) state $\ket\psi\in\K$, we can expand it on the normalized orthogonal basis $\ke{\ket{O_n}}_{n=0}^{L-1}$ as
\begin{align}\label{Expansion}
    \ket\psi=\sum_n\ket{O_n}\phi_n,\quad \phi_n=\avg{O_n|\psi}.   
\end{align}
Then we can define the Krylov complexity of the state $\ket\psi$ as
\begin{align}\label{KC1}
    K=J/P,
\end{align}
where
\begin{align}
    J=\sum_n n \abs{\phi_n}^2, \quad
    P=\sum_n \abs{\phi_n}^2.
\end{align}

\subsection{Krylov complexity of evolving states}

We consider the target state $\ket{\psi_\tau}$ generated by evolving the reference state $\ket0$ for time $\tau$ with $\L$, namely, \cite{Balasubramanian:2022tpr,Avdoshkin:2019euclidean,Balasubramanian:2022dnj}
\begin{align}
	\ket{\psi_\tau}=e^{-\tau \L} \ket{0},\quad \tau=\beta+it,\quad \beta, t\in\mathbb R,
\end{align}
and $\avg{\psi_\tau|\psi_\tau}=S(2\beta)$, where $S(\tau)=\bra0 e^{-\tau\L}\ket0$ is the {\it survival amplitude} for the state $\ket0$ to remain unchanged \cite{Balasubramanian:2022tpr}.   
We have introduced the inverse temperature $\beta$ and the real time $t$ to study the complexity due to imaginary and real time evolution. Note that the imaginary time evolution in this paper is different from the finite temperature construction of the Krylov basis in \cite{Balasubramanian:2022tpr}, where the authors change the Krylov basis at finite temperature.
Its expansion coefficient in \eqref{Expansion} is
\begin{align}\label{PsiToPhi}
	\phi_n(\tau)
	=\bra{0}\psi_n(\L)e^{-\tau\L}\ket{0}
	=\int_E \psi_n(E)e^{-\tau E}.
\end{align}
In particular, the survival amplitude is
$
	\phi_0(\tau)
	=\int_E e^{-\tau E}
    =S(\tau)
$.
If we regard $\phi_n(\tau)$ as a wave function at site $n$ on the Krylov chain, it will obey the Schr\"odinger equation following the recurrence relation \eqref{recurrence}
\begin{align}\label{SchIm}
	-\partial_\tau \phi_n(\tau)=b_{n+1}\phi_{n+1}(\tau)+a_n\phi_n(\tau)+b_n\phi_{n-1}(\tau),
\end{align}
with the initial condition 
$
    \phi_n(0)=\delta_{0n}
$.
The Krylov complexity \eqref{KC1} for the target state $\ket{\psi_\tau}$ is
\begin{align}
	K(t;\beta)=\frac{J(t;\beta)}{P(t;\beta)}
\end{align}
where 
\begin{align}
	&J(t;\beta)=\sum_n n \abs{\phi_n(\beta+it)}^2,\\
	&P(t;\beta)=\sum_n \abs{\phi_n(\beta+it)}^2
	=S(2\beta) \, .\label{ProbabilityConservation}
\end{align}
Since $P(t;\beta)$ is the total probability, it is conserved under real time $t$ evolution, which we indicate by defining $S(2\beta)$. In \eqref{ProbabilityConservation}, the (non-normalized) {\it transition probability} is defined as
\begin{align}
	\abs{\phi_n(\beta+it)}^2=\int_{E_1,E_2}e^{-\beta(E_1+E_2)-it(E_1-E_2)}\psi_n(E_1)\psi_n(E_2),
\end{align}
for the state $\ket0$ evolving to the Krylov state $\ket{O_n}$. In particular, the {\it survival probability} is given by the transition probability of $n=0$, $\abs{\phi_0(\tau)}^2=\abs{S(\tau)}^2$.

According to the Schr\"odinger equation \eqref{SchIm}, we have the imaginary time derivative
\begin{align}
	&\partial_\beta P(0;\beta)
	=\sum_n \kd{4 b_{n+1}\phi_{n+1}(\beta)\phi_{n}(\beta)-2a_n\phi_n(\beta)^2},\\
	&\partial_\beta J(0;\beta)
	=\sum_n \kd{ 2(2n+1) b_{n+1}\phi_{n+1}(\beta)\phi_{n}(\beta)-2na_n\phi_n(\beta)^2},
\end{align}
and the real time derivatives $\partial_t P(t;\beta)=0$ and
\begin{align}\label{Ehrenfest}
	&\partial_t^2 J(t;\beta)
	= 2\sum_{n=0}^{L-1} \kd{(b_{n+1}^2-b_n^2)\phi_n(\tau)\phi_n(\tau^*)-(a_{n+1}-a_n)b_{n+1}\phi_{(n+1}(\tau)\phi_{n)}(\tau^*)}, 
\end{align}
where $\tau=\beta+it$ and $b_0=0$ according to the convention in \eqref{Algorithm}. The r.h.s.~of \eqref{Ehrenfest} is just the expectation value of the commutator, such that we have
\begin{equation}
  \partial_t^2\bra{\psi_\tau}\hat K\ket{\psi_\tau}
  =-\bra{\psi_\tau}[[\hat K,\L],\L]\ket{\psi_\tau} \, .
\end{equation}
with Krylov complexity operator defined as $\hat K=\sum_n n\ket{O_n}\bra{O_n}$.
We refer to this equation as the {\it Ehrenfest theorem} of Krylov complexity, since it relates the second derivative of Krylov complexity to the expectation value of the gradient of the square of Lanczos coefficients. 
In this sense, it provides a classical equation of motion for the Krylov complexity.
This feature will become more clear in the continuum limit in App.~\ref{sec:Imaginary}. We note that a version of \eqref{Ehrenfest} for Krylov operator complexity in the case where $a_n=0$ was derived in \cite{Muck:2022xfc}. We will also explain the relation between the Ehrenfest theorem \eqref{Ehrenfest} and the ``complexity algebra'' \cite{Caputa:2021sib} in Sec.~\ref{sec:KCSFF}.

To calculate $K(t;\beta)$, we could start from $K(0;0)=0$, evolve it along imaginary time for $\beta$, get $K(0;\beta)$, evolve it along real time for $t$ with the initial condition $\partial_t K(t;\beta)|_{t=0}=0$, and finally get $K(t;\beta)$. 
Usually, we will integrate \eqref{Ehrenfest} over the real time and get the complexity difference $\Delta K(t;\beta)\equiv K(t;\beta)-K(0;\beta)=\int_0^t dt_2\int_0^{t_2} dt_1 \partial_{t_1}^2 J(t_1;\beta)/S(2\beta) $.
We discuss two limits of Krylov complexity below.

At the low temperature limit, we may use eigenstates with the lowest two energies $E_{0,1}$ to approximate the wave function
\begin{align}
	\phi_n(\beta+it)\approx \psi_n(E_0)e^{-(\beta+it) E_0} \abs{\avg{E_0|0}}^2+ \psi_n(E_1)e^{-(\beta+it) E_1} \abs{\avg{E_1|0}}^2.
\end{align}
The Krylov complexity converges to a constant plus an oscillation with frequency $E_{10}=E_1-E_0$, namely,
\begin{align}\label{Complexitylow}
	K_\text{low}(t;\beta)=\abs{\avg{E_0|0}}^2\sum_n n\psi_n(E_0)^2+ e^{-\beta E_{10}}2\cos(t E_{10})\abs{\avg{E_1|0}}^2\sum_n n\psi_n(E_0)\psi_n(E_1).
\end{align}

Similarly to the SFF \cite{Cotler:2016fpe},  the transition probability $\abs{\phi_n}^2$  is determined by the energy levels in the long-time average, where oscillating phases average to zero and terms with $E_p=E_q$ survive, namely
\begin{align}\label{probabilityAverage}
	(\abs{\phi_n(\beta+it)}^2)_\infty
	\equiv&\lim_{T\to\infty}\frac1T \int_0^T dt \sum_{pq}\abs{\avg{E_p|0}}^2\abs{\avg{E_q|0}}^2 e^{-\beta(E_p+E_q)+it(E_p-E_q)}\psi_n(E_p)\psi_n(E_q)\\
	=&\sum_p\abs{\avg{E_p|0}}^4 e^{-2\beta E_p}\psi_n(E_p)^2,
\end{align}
where we have assumed for simplicity that there is no degeneracy.
We refer to the time when $\abs{\phi_n}^2$ converges to this value as plateau time $t_p$. 
Then the long-time average of the complexity is given by
\begin{align}\label{ComplexityAverage}
	K_\infty(\beta)
	=\frac1{S(2\beta)}\sum_{pn}\abs{\avg{E_p|0}}^4 e^{-2\beta E_p}n\psi_n(E_p)^2.
\end{align}
The late time average at $\beta=0$ will be simplified if the reference state is taken to be a maximally entangled state in Sec.~\ref{sec:TFDRMT}.

\subsection{Continuum limit}\label{sec:ContinuumLimitSecond}

It is difficult to
solve the recurrence relation and  Schr\"odinger equation with general Lanczos coefficients on the discrete Krylov chain. To simplify this problem,
we consider the continuum limit $n\to x$, with $x$ a continuous coordinate, and solve the corresponding differential equations. 

\subsubsection{First-order formalism}\label{sec:ContinuumLimitFirst}

We may map the polynomials $\psi_n(E)$ and the wave function $\phi_n(\tau)$ to some continuous functions of $n$ (as given below in \eqref{continuumLimit} and \eqref{continuumLimitBackward}). Assuming that these functions depend smoothly on $n$, we may approximate their differences in $n$ by their derivative w.r.t.~$n$, and write the recurrence relation \eqref{recurrence} and Schr\"odinger equation \eqref{SchIm} as first-order differential equations \cite{Muck:2022xfc,Alishahiha:2022anw}. We refer to this approach as the first-order formalism of the continuum limit.
Based on this simplifying approach, we may easily derive the Krylov complexity in the continuum limit. However, we will see that the assumption of  smoothness is subtle and has to be clarified in a second-order formalism.

The authors of \cite{Muck:2022xfc,Alishahiha:2022anw} developed an approach to calculate the polynomial $\psi_n(E)$ and wave function $\phi_n(it)$ in the continuum limit as follows.
The continuum limit is defined to take the form
\begin{align}\label{continuumLimit}
	x_n=\epsilon n,\ \Psi(E,x_n)=i^n\psi_n(E),\ \Phi(t,x_n)=i^n\phi_n(\beta+it), \ b(x_n)=b_n,\ a(x_n)=a_n,
\end{align}
which is valid when $a_n$, $b_n$, $i^n\psi_n(E)$, and $i^n\phi_n(\beta+it)$ are smooth functions of $n$. For real time $t$,
the recurrence relation \eqref{recurrence} and the Schr\"odinger equations \eqref{SchIm}  become
\begin{align}
    i(E-a)\Psi=&~\epsilon  b'\Psi+2\epsilon b \Psi' +O(\epsilon^2),\label{CSchPsi}\\
    -(\partial_t+ia) \Phi=&~\epsilon b' \Phi + 2\epsilon b\Phi'  + O(\epsilon^2)\label{CSchRe},
\end{align}
where $b'=\partial_xb$, and equivalently for the other variables. The above two equations are related by the transformation \eqref{PsiToPhi} from energy $E$ to time $t$.
Due to the $\epsilon b' \Psi$ term in \eqref{CSchPsi}, the norm in \eqref{Orthogonal} is not preserved by the evolution along $x$.
Using the coordinate $y$ with $dy=dx/(2\epsilon b(x))$ and with the new variable $\tilde \Psi=\sqrt{b}\Psi$ and $\tilde \Phi=\sqrt{b}\Phi$, these equations simplify to  \begin{align}
    (-iE+ia+\partial_y)\tilde\Psi=(\partial_t+ia+\partial_y)\tilde\Phi=0.
\end{align}
Using the coordinate  $x$, the solutions then become
\begin{align}
    \Psi(E,x)=\sqrt{\frac{b(0)}{b(x)}}\Psi(E,0)\exp\kc{i\int_0^x \frac{E-a(x')}{2\epsilon b(x')}dx'},\label{PolynomialC}\\
    \Phi(t,x)=\frac1{\sqrt{b(x)}}f(t_-(t,x))\exp\kc{-i\int_0^x \frac{a(x')}{2\epsilon b(x')} dx'}, \label{WaveFunctionC}
\end{align}
where the function $f(t_-)$ is determined by the initial condition and $t_-(t,x)$ labels the characteristic curves \cite{Muck:2022xfc}
\begin{align}\label{CharacteristicCurvesForward}
	t_-(t,x)=t-\int^x \frac{dr}{2\epsilon b(r)}.
\end{align}
This shows that the wave function $\Phi(t,x)$ propagates forward with a local velocity $2\epsilon b(x)$,  from $n=0$ to $n=L$.
Notice that \eqref{PolynomialC} and \eqref{WaveFunctionC} do not contain the end point of the Krylov chain $x=\epsilon L$ since the last state in the Krylov basis \eqref{KrylovBasis} is $\ket{O_{L-1}}$.
Since the initial condition $\phi_n(0)=\delta_{0n}$ is highly discontinuous, the continuum limit is valid only when the wave function spreads out.

However, we notice that the discrete  recurrence relation \eqref{recurrence} and  Schr\"odinger \eqref{SchIm} enjoy the parity symmetry  $n\to L-n$, but their continuum versions \eqref{CSchPsi} and \eqref{CSchRe} break the parity $x\to \epsilon L-x$. As a result, the characteristic curves \eqref{CharacteristicCurvesBackward} have a preferred direction. The breaking of parity is due to the assumption on the smoothness of $i^n\psi_n(E)$ and $i^n\phi_n(\beta+it)$ as functions of $n$ in the continuum limit. If we consider an alternative continuum limit, namely
\begin{align}\label{continuumLimitBackward}
    \Psi(E,x_n)=i^{-n}\psi_n(E),\quad \Phi(t,x_n)=i^{-n}\phi_n(\beta+it),
\end{align}
we find 
\begin{align}
    -i(E-a)\Psi=&~\epsilon  b'\Psi+2\epsilon b  \Psi' +O(\epsilon^2),\\
    (\partial_t+ia) \Phi=&~\epsilon b' \Phi + 2\epsilon b\Phi'     + O(\epsilon^2).
\end{align}
This result corresponds to the backward characteristic curves
\begin{align}\label{CharacteristicCurvesBackward}
    t_+(t,x)=t+\int^x \frac{dr}{2\epsilon b(r)}.
\end{align}
This backward propagation will be important after the wave function is reflected by the endpoint at $n=L$. 
The forward propagation and backward propagation are unified by the second-order formalism presented in the next section. 

Finally, we note that the polynomials $\psi_n(E)$ obtained from \eqref{PolynomialC} have some artefacts.
First, as a function of $E$, $\psi_n(E)$ is a Fourier mode of frequency $L(1-\sqrt{1-n/L})$ instead of a polynomial of degree $n$. 
Second, in general it does not obey the orthogonality and completeness relations \eqref{Orthogonal}\eqref{Complete} and is not normalized to $1$. 
Third, it usually takes a complex value. The first two artefacts are the results of continuum limit. 
Finally, since \eqref{CSchPsi} does not preserve the normalization, we have to renormalize $\psi_n(E)$ for each $n$ and $E$. 
The third artefact is solved by the second-order formalism as well.

\subsubsection{Second-order formalism}

We will adopt the following second-order formalism, which develops from the approach in \cite{Muck:2022xfc}.
Applying the recurrence relation \eqref{recurrence} and Schr\"odinger equations \eqref{SchIm} twice, we obtain 
\begin{align}\label{recurrence2}
    E^2 \psi_n
    =&~\sum_{ml} L_{nm}L_{ml}\psi_l
    =c_{n+1}\psi_{n+2}+d_{n+1}\psi_{n+1}+e_n\psi_n+d_{n}\psi_{n-1}+c_{n-1}\psi_{n-2},\\
    \label{Sch2}
    -\partial_t^2 \phi_n
    =&~\sum_{ml} L_{nm}L_{ml}\phi_l
    =c_{n+1}\phi_{n+2}+d_{n+1}\phi_{n+1}+e_n\phi_n+d_{n}\phi_{n-1}+c_{n-1}\phi_{n-2},
\end{align}
where we dropped the arguments of $\psi_n(E)$ and $\phi_n(\tau)$ and the coefficients are
\begin{align}
    c_n=b_nb_{n+1},\quad 
    d_n=b_{n}(a_{n-1}+a_{n}),\quad
    e_n=b_n^2+a_n^2+b_{n+1}^2.
\end{align}
When $a_n=0$ and then $d_n=0$, the second-order formalism results \eqref{recurrence2} and \eqref{Sch2} are factorized into even and odd parts, respectively \cite{Muck:2022xfc}. This happens in the case of even-parity spectrum  $\{ E_p\} = \{-E_p\}$.
Moreover, we consider even $L$ for simplicity. 
The recurrences of even sector $\ke{\psi_0,\psi_2,\cdots,\psi_{L-2}}$ and odd sector $\ke{\psi_1,\psi_3,\cdots,\psi_{L-1}}$ are decoupled. 
The same applies to the evolution of wave function. 
It is therefore not appropriate in general to assume that the even sector smoothly connects to the odd sector. 
We therefore proceed as follows. We consider the continuum limit
\begin{align}\label{continuumLimit2}\begin{split}
x_n=\epsilon n,&\quad 
\Psi(E,x_n)=i^n\psi_n(E),\quad \Phi(t,x_n)=i^n\phi_n(\beta+it), \\ 
 &c(x_n)=c_n,\quad 
 g(x_n)=e_n-2c_n,
 \end{split}\end{align}
where we assume that the $i^n\psi_n(E)$ and $i^n\phi_n(\beta+it)$ for even $n$ and for odd $n$ are continuous respectively. The second-order formalism of \eqref{recurrence2} and \eqref{Sch2}  becomes, for either the even sector or the odd sector,
\begin{align}
    (-E^2+g)\Psi
    =\epsilon^2 (4 c'\Psi'+c''\Psi+4c\Psi'')+ O(\epsilon^4), \\
    (\partial_t^2+g)\Phi 
    =\epsilon^2 (4 c'\Phi'+c''\Phi+4c\Phi'')+ O(\epsilon^4).
\end{align}
Since these are real equations, we may obtain real solutions with real boundary conditions.
Using the coordinate $y$ with $dy=dx/(2\epsilon \sqrt{c(x)})$ and with the variables $\tilde \Psi=c^{1/4}\Psi,\tilde \Phi=c^{1/4}\Phi$, these equations are simplified into two wave equations
\begin{align}\label{WaveEq}
\left(-E^2-\partial_y^2+V\right)\tilde\Psi
=\left(\partial_t^2-\partial_y^2+V\right)\tilde\Phi
=0 \, ,
\end{align}
with potential $V=g-(\frac14\partial_y\ln c)^2$. 
Thus, the characteristic curves are
\begin{align}\label{CharacteristicCurves}
    t_\pm(t,x)=t\pm \int^x \frac{dx'}{2\epsilon \sqrt{c(x')}} \, .
\end{align}
They correspond to the forward and backward characteristic curves also given by the first-order formalism, \eqref{CharacteristicCurvesForward} and \eqref{CharacteristicCurvesBackward}. 
Since $\psi_0(E)=1$ in the even sector and $\psi_1(E)=E/b_1$ in the odd sector, we impose the boundary conditions
\begin{align}\label{BoundaryCondition}
    \text{even}\ n:&\quad \Psi(E,0)=1,\quad \Psi^{(0,1)}(E,0)=0
    ,\\
    \text{odd}\ n:&\quad \Psi(E,0)=0,\quad \Psi^{(0,1)}(E,0)=iE/b_1.
\end{align}
for simplicity of the solutions.
The functions $\Psi(E,x)$ determined by solving the wave equations with the corresponding boundary conditions
for the even sector and odd sector, respectively. A better approximation may be obtained by modifying the boundary conditions according to the values of $\psi_2(E)$ and $\psi_3(E)$.

\subsection{Krylov complexity for the TFD state}\label{sec:TFDRMT}

The Krylov approach relies on the choice of the Liouvillian and reference state. 
In this subsection, to study the direct relation between chaos in the spectrum of a Hamiltonian $H$ and Krylov state complexity, we will construct the Liouvillian from a Hamiltonian $H$ and consider a maximally entangled state as the reference state. This is motivated by the construction of Ref.~\cite{Balasubramanian:2022tpr}. 

Consider a Hilbert space $\H$ with dimension $D=\dim \H$, and a Hamiltonian $H$ with eigenstates $\ket{E_i}_1, \ i=0,1,2,\cdots,D-1$, where the subscript ``$_1$'' denotes the single Hilbert space.
We consider a double-copy of the Hilbert space $\H_L\otimes\H_R$. 
Given a copy of energy basis $\ket{E_i,E_j}=\ket{E_i}_1\otimes\ket{E_j}_1$, we can define a maximally entangled state in the double-copied Hilbert space 
\begin{align}\label{MaxEntangle}
	\ket{0}=\frac1{\sqrt{D}}U_L\otimes U_R \sum_{i=0}^{D-1}\ket{E_i,E_i}.
\end{align}
where $U_{L,R}$ is an unitary operator acting on $\H_{L,R}$.
We will take $\ket0$ as the reference state.
We consider the Liouvillian given by
\begin{align}\label{Liouvillian}
    \L=H\otimes\I,
\end{align}
where $\I$ is the identity operator. Since the choice of unitary operators in \eqref{MaxEntangle} will not affect the moment $\bra0\L^j\ket0$, we are free to chose $U_L=U_R=\I$. The Lanczos algorithm only depends on the spectrum of $H$.

The Krylov space $\K$ is spanned by $\ke{\L^n\ket0=(1/\sqrt{D})\sum_i E_i^n\ket{E_i,E_i}}$, which is a subspace of the space of equal-energy states $\H_{\rm eq}=\ke{\ket{E_i,E_i}}_{i=0}^{D-1}$. Their inner product can be written as the trace $\Tr$ in the single-copy Hilbert space,
\begin{align}\label{Inner}
	\bra0 f(\L)\ket0
	=\frac1D\sum_{i=0}^{D-1} f(E_i)=\frac1D \Tr f(H).
\end{align}

Now the moment is $\mu_{j}=\Tr[H^{j}]/D$, and especially, $b_1^2=\bra0 \L^2\ket0=\Tr[H^2]/D$. 
Notice that $\ke{E_i^n}$ for $0\leq i,n<D$ form a Vandermonde matrix, whose rank is the dimension of Krylov space $L=\dim\K$. $L$ is reduced by the degree of degeneracy in the spectrum for the following reasons. Consider the decomposition $H_{\rm eq}=\bigoplus_{p=0}^{l-1} \H_p$ where $\H_p$ is a subspace with $m_p$-fold degeneracy $E_{p,0}=E_{p,1}=\cdots=E_{p,m_p-1}$. For each $\H_p$, we may construct a basis where the first state is $\ket{E_p}=(1/\sqrt{m_p})\sum_{k=0}^{m_p-1}\ket{E_{p,k},E_{p,k}}$, which overlaps with the reference state as $\avg{E_p|0}=\sqrt{m_p/D}$, and the remaining $m_p-1$ states are orthogonal to $\ket0$. 
Then the dimension of Krylov space is reduced by $m_p-1$ for each $\H_p$. 
Since the Vandermonde matrix of the non-degenerate spectrum $\ke{E_p^n}$ for $0\leq p,n<l$ has full rank, $L$ equals the number of the non-degenerate energies, namely $L=l$. 
In this paper, we focus on the case that the spectrum is uniformly $d$-fold degenerate. 
So, the Krylov space of the maximally entangled state has dimension $L=D/d$ and overlap $\abs{\avg{E_p|0}}^2=1/L$. Obviously, $\L_\K$ is diagonal on the basis $\ket{E_p}$. 
Notice the difference between the spectral density $\rho(E)$ of $\L_\K$ in \eqref{Density} and the spectral density of $H$.
The Krylov basis defined in \eqref{KrylovBasis} obeys the orthogonality and completeness relations \eqref{Orthogonal}\eqref{Complete}, with \eqref{Measure} equal to \eqref{Inner}.

The target state 
\begin{align}
    \ket{\psi_\tau}
    =e^{-\tau\L}\ket0
	=\frac1{\sqrt{D}}\sum_{i=0}^{D-1}e^{-\tau E_i}\ket{E_i,E_i}
\end{align}
with $\tau=\beta+it$ is a TFD state with inverse temperature $2\beta$ and evolved by the left side Hamiltonian for time $t$.
Its coefficient on the Krylov basis, {\it i.e.} the wave function on the Krylov chain, is
\begin{align}\label{WaveFunctionTFD}
	\phi_n(\tau)
 =\frac1D\Tr[\psi_n(H)e^{-\tau H}].
\end{align}
Obviously, the survival probability is related to the SFF \eqref{SFF} as
$
	\abs{\phi_0(\tau)}^2
	=(1/D^2) \abs{Z(\tau)}^2.
$
With $S(2\beta)=Z(2\beta)/D$, the Krylov complexity is written as
\begin{align}\label{KCTFD}
	K(t;\beta)=\frac{\sum_n n\abs{\Tr[\psi_n(H)e^{-(\beta+it) H}]}^2}{D\Tr[e^{-2\beta H}]}.
\end{align}
Thus, the Krylov complexity for the TFD state only depends on the spectrum.

The above Krylov approach based on the maximally entangled state works at finite $D$. In the $D\to\infty$ limit, one have to firstly regularize the dimension $D$ by truncating the spectrum. Then rescale the spectrum into an finite energy interval such that all the moments $\mu_j$ for finite $j$ are finite. Finally, one can take the $D\to\infty$ limit. Taking the Gaussian ensembles as examples, we will normalize the Hamiltonian so that the spectral density $\avg{\rho(E)}$ in ensemble average vanishes when $E\not\in[-2,2]$ at $D\to\infty$.

The Krylov space for the maximally entangled state has overlap $\abs{\avg{E_0|0}}^2=1/L$. At $\beta=0$, the long-time averages of the transition probability and of the Krylov complexity, given by \eqref{probabilityAverage} and \eqref{ComplexityAverage}, are simplified due to the orthogonality relation \eqref{Orthogonal}, as shown in \cite{Rabinovici:2020operator,Rabinovici:2022beu}:
\begin{align}\label{AverageTFD}
    (\abs{\phi_n}^2)_\infty=\frac1L,\quad K_\infty=\frac {L-1}2.
\end{align}
These results are independent of whether chaotic behavior is present or not. They are essentially a consequence of taking the maximally entangled state as the reference state. However, we will show in Sec.~\ref{sec:firstlook} that the fluctuations of transition probability and Krylov complexity are indeed sensitive to the presence of chaotic behavior.

\section{Krylov complexity at early times}\label{sec:Earlytime}

Here, we will consider the Hamiltonian $H$ drawn from the Gaussian orthogonal ensemble (GOE),  the Gaussian unitary ensemble (GUE) and the Gaussian symplectic ensemble (GSE).
They belong to the $\tilde\beta$-Hermite (Gaussian) ensemble with Dyson index $\tilde\beta=1,2,4$ respectively \cite{Dumitriu:2002beta}.
The measures of their random spectra are given by \eqref{MeasureBeta} in App.~\ref{sec:PolynomialEnsemble}.
Due to the level repulsion in RMT, the spectra in both the GOE and GUE are non-degenerate, where $L=D$, and the spectrum in the GSE is doubly degenerate, where $L=D/2$. 
To simplify the notation, we are free to rescale the Hamiltonian so that the first Lanczos coefficient $b_1=1$. 
To recover the dimension, we can rescale the energies and times as $E_p\to E_p/\lambda,\ a_n\to a_n/\lambda, \ b_n\to b_n/\lambda$, and $\tau\to \tau \lambda$, where $\lambda$ has the dimension of energy.

\begin{figure}
    \centering
    \includegraphics[height=0.35\linewidth]{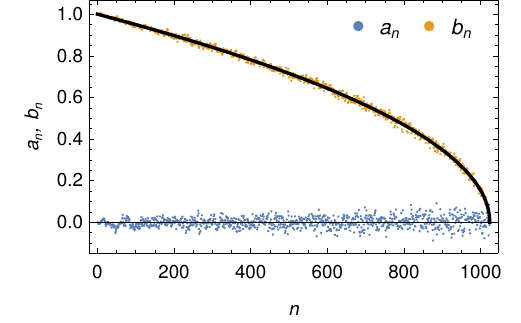}
    \caption{The Lanczos coefficients of a maximally entangled state evolving with the Liouvillian $\L=H\otimes\I$ where the $H$ is taken from the GUE with dimension $D=4096$. The black curve is a $\sqrt{1-n/D}$.}
    \label{fig:LanczosGUETFD}
\end{figure}

In Fig.~\ref{fig:LanczosGUETFD}, we display the Lanczos coefficients for the Krylov space of maximally entangled states for a realization of the GUE. 
When $n$ increases, $a_n$ fluctuates around $0$ and $b_n$ decreases from $1$ to $0$. 
Their expectation values in the large $L$ limit are further discussed in Sec.~\ref{sec:AverageLanczos}.
Their fluctuations become stronger when $n$ increases.

\begin{figure}
    \centering
    \includegraphics[height=0.35\linewidth]{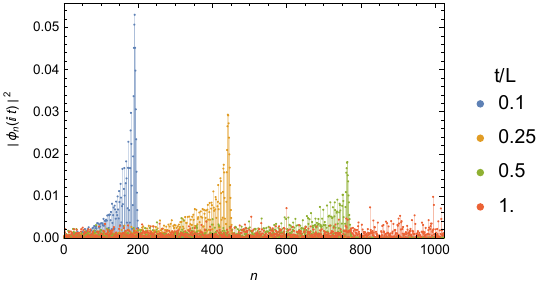}
    \includegraphics[height=0.35\linewidth]{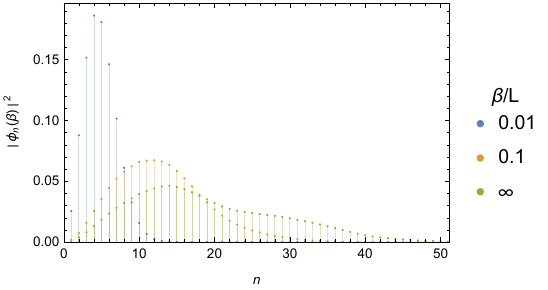}
    \caption{The snapshots of the transition probability $\abs{\phi_n(it)}^2$ and $\abs{\phi_n(\beta)}^2$ on the Krylov chain in one realization of the GUE with $L=1024$.}
    \label{fig:wave}
\end{figure}

In Fig.~\ref{fig:wave}, we show the snapshots of the transition probability at some instants of real time $t$ and imaginary time $\beta$ in one realization of the GUE. 
Along the real time evolution, the profile consists of a shock wave, a long tail, and some residual noise following. 
The strength of the shock wave and the tail decreases along the time, but the tail becomes longer. 
The shock wave becomes tiny when reaching the end of the Krylov chain. 
Along the imaginary time evolution, the wave function is localized at the ground state of the Schr\"odinger equation \eqref{SchIm}, which is similar to the fate of the wave function in the Krylov space of open systems \cite{Liu:2022god,Bhattacharya:2022gbz,Bhattacharjee:2022lzy,Bhattacharya:2023zqt}.

In this section, we will develop an analytical approach for obtaining the evolution of Krylov complexity at early times.

\subsection{Lanczos coefficients in the large dimension limit}\label{sec:AverageLanczos}

For the Liouvillian drawn from the $\tilde\beta$-Hermite ensemble and a fixed reference state $\ket{0'}$, the recent results of \cite{Balasubramanian:2022dnj} show that the Lanczos coefficients obey the statistical distribution function
\begin{align}
    \rho(a_n)=\frac1{\sqrt{\tilde\beta L}}N(0,2)(a_n),\quad
    \rho(b_n)=\frac1{\sqrt{\tilde\beta L}}\chi_{(L-n)\tilde\beta} (b_n),
\end{align}
where $N(r, s)$ is the Gaussian distribution with mean $r$ and variance $s$, and
$\chi_r(x)$ is the chi-distribution, given by $\chi_{r}(x) =2^{-r/2} x^{r-1} e^{-x^2/2}/\Gamma(r/2)$.  The Lanczos coefficients have the expectation value and variance  \cite{Balasubramanian:2022dnj}
\begin{align}\label{LanczosStatistic}
    \avg{a_n}=0,\quad 
    \Delta^2(a_n)=\frac{4}{\tilde\beta L},\quad
    \avg{b_n^2}=1-\frac{n}{L},\quad
    \Delta^2(b_n^2)=\frac{2}{\tilde\beta L}\kc{1-\frac nL},\\
    \avg{b_n}=\sqrt{\frac{2}{\tilde\beta L}}\frac{\Gamma \kc{(L-n)\tilde\beta/2+1/2}}{ \Gamma \kc{(L-n)\tilde\beta/2}},\quad 
    \Delta^2(b_n)=1-\frac{n}{L}-\avg{b_n}^2,
\end{align}
where $\avg{\cdots}$ denotes average over the $\tilde\beta$-Hermite ensemble, and $\Delta^2$ denotes the variance. 
In the large $L$ limit with finite $x=n/L$, we have $\Delta^2(b_n)=O(L^{-1})$. So the variance is small compared to the average of $b_n$. 
Thus, in the large $L$ limit, we can take the expectation values
\begin{align}\label{LanczosLimit}
    a_n=0,\quad b_n =\sqrt{1-\frac nL}, 
\end{align}
where the deviation from the convention $b_1^2=1$ is negligible for large $L$. In our analysis below, we will maintain \eqref{LanczosLimit} for conciseness. The same scaling behavior in \eqref{LanczosLimit} was applied to the Krylov operator complexity at late times in \cite{Kar:2021nbm}.

We notice that, for the maximally entangled state, the Liouvillian $\L$ in \eqref{Liouvillian} is factorized and thus it is not a random Hamiltonian acting on the double-copy Hilbert space $\H_L\otimes\H_R$.  
The statistics of Lanczos coefficients \eqref{LanczosStatistic} is not simply applicable. 
However, we find that if we take the limit \eqref{LanczosLimit} as the tridiagonalization of $H$ and apply the algorithm \eqref{Algorithm} on the maximally entangled state $\ket0$ with the Liouvillian $\L$, we get the same Lanczos coefficients \eqref{LanczosLimit}. 
In order words, the limit \eqref{LanczosLimit} is a fixed point of the algorithm on maximally entangled state \eqref{MaxEntangle}. 
So, \eqref{LanczosLimit} may be taken as an approximation for the Lanczos coefficients of maximally entangled state in the large $L$ limit, as shown in Fig.~\ref{fig:LanczosGUETFD}.

We now consider the continuum limit \eqref{continuumLimit} of the Lanczos coefficients \eqref{LanczosLimit}, which is given by $a(x)=0,\, b(x)=\sqrt{1-x/(\epsilon L)}$. Then the $y$ coordinate in the continuum limit is $y(x)=L \left(1-\sqrt{1-x/(L \epsilon) }\right)$. From the characteristic curves \eqref{CharacteristicCurves}, at time $t$ the shock wave will reach the site
\begin{align}\label{shock}
    n(t)=t\kc{2-\frac tL},\quad 0<t\leq 2L.
\end{align}
Thus the shock wave reaches the last site at $t=L$. When $t>L$, the shock wave gets reflected and travels backward.

The authors of \cite{Balasubramanian:2022dnj} gave an approximate way to relate the density of state to the Lanczos coefficients in the large $L$ limit. 
Setting $b_n=0$ for a relatively small number of $n$, {\it e.g.}, $n=m s$ for $m=1,2,...,r$ with integers $r,s\sim\sqrt{L}$ and $L=rs$, the density of states is slightly affected, however we neglect this effect since only a small proportion of $b_n$ are sent to zero for large $L$.
This approach makes the tridiagonal matrix $\mathbf L$ into a block diagonal matrix with $r$ blocks of size $s\times s$. Furthermore, in the large $L$ limit, the Lanczos coefficients change smoothly, {\it e.g.} \eqref{LanczosLimit}. 
We can further approximate the $a_n$'s and $b_n$'s in the $m$-th block by their mean values $\bar a_m,\bar b_m$. 
From the case of constant Lanczos coefficients \eqref{ConstantDensity}, the density of state in the $m$-th block is
\begin{align}\label{BlockDensity}
    \rho_m(E)=\frac Ls\frac{\Theta(4\bar b_m^2-(E-\bar a_m)^2)}{\pi \sqrt{4\bar b_m^2-(E-\bar a_m)^2}},
\end{align}
with normalization $L/s$. 
The total density of state is
\begin{align}\label{LanczosToDensity}
    \rho(E)=\sum_m \rho_m(E) 
    \approx \frac{L}{\pi}\int_0^1 dx\frac{\Theta(4 b(x)^2-(E-a(x))^2)}{\sqrt{4 b(x)^2-(E-a(x))^2}}
\end{align}
where $a_n=a(x),\, b_n=b(x)$ are introduced in the continuum limit with $x=n/L$. 
Obviously, via \eqref{LanczosToDensity}, in the large $L$ limit the Lanczos coefficients \eqref{LanczosLimit} give the semi-circle law 
\begin{align}\label{SemiCircle}
	\rho_{\rm sc}(E)=\frac L{2 \pi }\sqrt{4-E^2}.
\end{align}
With the above ingredients, in the next subsection we will find an analytical approximation to the Krylov complexity at early times.

\subsection{Krylov complexity from spectral form factor}\label{sec:KCSFF}

Based on the Ehrenfest theorem \eqref{Ehrenfest} and the limit value of Lanczos coefficients \eqref{LanczosLimit} in the RMT of $\tilde\beta$-ensemble, we now give a direct relation between the Krylov complexity for maximally entangled states and the SFF at early times.

To proceed, we insert \eqref{LanczosLimit} into the Ehrenfest theorem \eqref{Ehrenfest} and intermediately find the simply combinations of coefficients 
\begin{align}\label{LanczosDifference}
    b_{n+1}^2-b_n^2=\delta_{n0}-1/L,
    \quad
    (a_{n+1}-a_n)b_{n+1}=0.
\end{align}
It is reminiscent of the ``complexity algebra'' in \cite{Caputa:2021sib,Hornedal:2022pkc}. In their case, $b_{n+1}^2-b_n^2=An+B,\ \forall n\geq0$ and then $\ke{\L,\mathcal B, \tilde K}$ form a closed algebra, where $\mathcal B=[\L,\hat K]$, $\tilde K=A\hat K+B$.
The closed algebra completely determines the evolution of complexity. 
However, here we have an ``anomaly'' $\delta_{n0}$, which prevents the ``complexity algebra'' from being closed. 
From the $n=0$ term in the Ehrenfest theorem \eqref{Ehrenfest}, we obtain that the external input for the evolution of complexity is $\abs{\phi_0(\tau)}^2$, which is proportional to the SFF according to \eqref{WaveFunctionTFD}. 
More precisely, we get the following expression of the second derivative of Krylov complexity,
\begin{align}\label{EhrenfestTFD}
    &\partial_t^2 K(t;\beta)
	\approx 
 \frac2D\frac{|Z(\beta+it)|^2}{Z(2\beta)}-\frac2L
	=\frac2{DZ(2\beta)}\sum_{p\neq q} e^{-\beta(E_p+E_q)-it(E_p-E_q)},
\end{align}
where in the last expression, only different levels will contribute to the summation.
This equation states that under the approximation for the Lanczos coefficient \eqref{LanczosLimit}, the second derivative of Krylov is given by the SFF \eqref{SFF}. 
In deriving this equation, we should note that the wave function also implicitly depends on the RMT Hamiltonian.
Thus, we have actually neglected the statistical correlation between the Lanczos coefficients and the wave function resulted from the RMT.

Surprisingly, if we take the double integral over the time on both sides of \eqref{EhrenfestTFD}, we obtain exactly the {\it spectral complexity} defined in \cite{Iliesiu:2021ari},
\begin{align}\label{SpectralComplexity}
    C(t;\beta)=\frac{1}{DZ(2\beta)}\sum_{p\neq q}\kd{\frac{\sin(t(E_p-E_q)/2)}{(E_p-E_q)/2}}^2 e^{-\beta(E_p+E_q)}.
\end{align}
This implies that the Krylov state complexity and the spectral complexity are related by the Ehrenfest theorem in Krylov space. 
The authors of \cite{Alishahiha:2022anw} also proposed an equivalence between Krylov complexity and spectral complexity in the continuum limit. 
However, since we neglect the fluctuation of the Lanczos coefficients, we will see that the two complexities match each other through the linear growth region but deviate from each other at late times. 
We will distinguish the Krylov complexity $K$ and its approximation from the Ehrenfest theorem, {\it i.e.} the spectral complexity $C$.

Notice that, beyond the plateau time, the SFF converges to the plateau value $\abs{Z(\beta+it)}^2\to d Z(2\beta)$ in ensemble average \cite{Cotler:2016fpe} with $d$-fold degenerate spectrum. 
Then \eqref{EhrenfestTFD} will vanish, which is a necessary condition for the saturation of spectral complexity $C$. 
However, it is not a sufficient condition, as \eqref{EhrenfestTFD} does not ensure that $\partial_t C(t;\beta)$ vanishes. It is necessary to check the saturation case by case. 
In Ref.~\cite{Iliesiu:2021ari}, the authors calculated the spectral complexity in the microcanonical ensemble for the GUE, GOE, GSE. Here we will work on the canonical ensemble, focus on the GUE in the main text and leave the calculation of the GOE and GSE in App.~\ref{sec:SpectralComplexityGaussian}. 
Finally, we will compare the numerical result of Krylov complexities to those spectral complexities.

For the GUE, the one-point function of the spectral density $\avg{\rho(E)}$ at large $L$ obeys the semicircle law \eqref{SemiCircle}, where the bracket is the matrix ensemble average. 
Its two-point correlation is given by the sine kernel \cite{Cotler:2016fpe,Cotler:2017jue,Liu:2018hlr}
\begin{align}\label{sine}
	\avg{\rho(E_1)\rho(E_2)} =\avg{\rho(E)}\delta(s)+\avg{\rho(E_1)}\avg{\rho(E_2)} \kd{1-\frac{\sin^2\kc{\pi\avg{\rho(E)}s}}{\kc{\pi\avg{\rho(E)}s}^2}},
\end{align}
where $s=E_1-E_2$ and $E=(E_1+E_2)/2$. 
The sine kernel shows the short-range correlation between the spectrum. 
By Fourier transformation, we obtain the SFF 
\begin{align}\label{SFFEnergy}
	\avg{\abs{Z(\beta+it)}^2}
 =\abs{\avg{Z(\beta+it)}}^2+\int dE e^{-2\beta E}\min\ke{\frac t{2\pi},\avg{\rho(E)}},
\end{align}
whose first (second) term is the disconnected (connected) part. 
From \eqref{SemiCircle}, the disconnected part is $\abs{\avg{Z(\beta+it)}}^2=\frac{L^2}{\beta ^2+t^2}\abs{I_1(2\beta+2i t)}^2$, which contributes the slope of SFF. 
The connected part contributes to the ramp of SFF via the $t/2\pi$ in the integrand in the energy window satisfying $t<2\pi\avg{\rho(E)}$ and contributes to the plateau when the energy window shrinks to zero.

We were unable to find an analytic formula of the SFF for a general $\beta$ \cite{Liu:2018hlr}. 
So, we will consider two limits $\beta=0$ and $\beta\gg1$ below.

\subsubsection{Infinite temperature limit}

We first consider the infinite temperature limit $\beta=0$ so that the SFF and Krylov complexity are sensitive to the full spectrum. 
Evaluating the integral \eqref{SFFEnergy} with \eqref{SemiCircle}, we obtain the SFF  \cite{Brezin:1997SFFRMT,Liu:2018hlr}
\begin{align}
	\avg{\abs{Z(it)}^2}=\Re\kd{\frac{L^2 J_1(2 t)^2}{t^2}+\frac{t}{\pi}\sqrt{1-\frac{t^2}{4L^2}}+\frac{2 L}{\pi}\cos ^{-1}\left(\sqrt{1-\frac{t^2}{4L^2}}\right)},
\end{align}
where $J_n$ is the $n$-th Bessel function of the first kind. 
Inserting this into the Ehrenfest theorem \eqref{EhrenfestTFD} and taking the double-time integral with the initial condition $C^{(1,0)}(0;0)=C(0;0)=0$, we obtain the spectral complexity
\begin{align}\label{KGUETFD}
	C(t;0)
 =&~
 \, _1F_2\left(-\frac{1}{2};1,2;-4 t^2\right)-1+L-\frac{16 t}{3 \pi }\\
 &~+\Re\kd{\frac {t}{6\pi}  \kc{\frac{t^2}{L^2} +26} \sqrt{1-\frac{t^2}{4L^2} }-\frac {2L}{\pi} \left(\frac{t^2}{L^2}+1\right) \sin ^{-1}\left(\sqrt{1-\frac{t^2}{4L^2}}\right)},\nn
\end{align}
which asymptotes to a constant at $t\to\infty$. 
More precisely, the spectral complexity has the asymptotic behavior
\begin{align}\label{SpectralComplexityGUE}
	C(t;0)\to&
	\begin{cases}
		t^2, & t\ll 1\\
		\frac{16}{3 \pi }t, & 1\ll t\ll L\\
		L,	& L\ll t
	\end{cases}
\end{align} 
The comparison between our analytical expression and the numerical calculation for the Krylov complexity at $\beta = 0$ is shown in the right panel of Fig.~\ref{fig:SFFKCGUETFDbeta}. 
They agree well before the peak time of the Krylov complexity. 
The linear growth $16t/{3\pi}$ is the result of the double-time integral of the slope of the SFF, which is uniquely determined by the one-point function of the spectral density and not the spectral correlation. 
In Sec.~\ref{sec:nonChaotic}, we will show that an uncorrelated spectrum leads to the quadratic-to-linear growth in complexities.
Similarly, in App.~\ref{sec:ConstantLanczos}, we calculate the complexities in the case of constant Lanczos coefficients and find the same quadratic-to-linear growth. We therefore state that the linear growth of the Krylov complexity with a maximally entangled reference state is not related to chaos.
The saturation value of the spectral complexity $C(\infty;0)=L-1$ is different from the saturation value of the Krylov complexity $K_\infty=(L-1)/2$ in \eqref{AverageTFD}, as shown in the right panel of Fig.~\ref{fig:SFFKCGUETFDbeta}.
The saturation of the spectral complexity after the plateau time is due to the cancellation between the factor $\sin^2\kc{\pi\avg{\rho(E)}x}$ in the sine kernel and the Fourier mode $\sin^2\kc{tx/2}$ in \eqref{SpectralComplexity} at either the imaginary infinity $x\to i\infty$ or $x\to -i\infty$. This saturation is related to chaos. Meanwhile, the saturation of the Krylov complexity for maximally entangled states is due to the discrete spectrum, which is not necessarily related to chaos.

\begin{figure}
    \centering
    \includegraphics[height=0.35\linewidth]{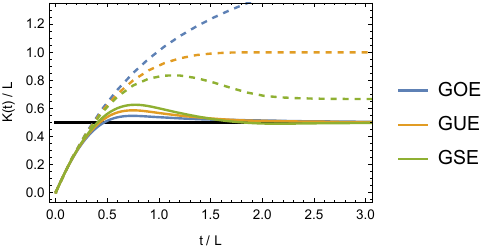}
    \caption{The Krylov complexities $K(t;0)$ (solid curves) of maximally entangled state and spectral complexities $C(t;0)$ (dashed curves) as functions of time $t$ in the GOE, GUE, and GSE. The Krylov complexities are calculated numerically from the $\tilde\beta$-ensemble with $L=256$ and $128$ realizations. The spectral complexities are derived from \eqref{SpectralIntegral}.}
    \label{fig:KCEns}
\end{figure}

In App.~\ref{sec:SpectralComplexityGaussian}, the spectral complexities in the GOE and GSE exhibit the same quadratic-to-linear growth as in \eqref{SpectralComplexityGUE}. But they saturate to $(L/2) \log t$ and $2L/3$ respectively when $t\gg L$. 
In Fig.~\ref{fig:KCEns}, we compare the numerical Krylov complexities and the spectral complexities for the three Gaussian ensembles, where they agree with each other at early times only.

\subsubsection{Low temperature limit}\label{sec:LowTemperature}

\begin{figure}
    \centering
    \includegraphics[height=0.3\linewidth]{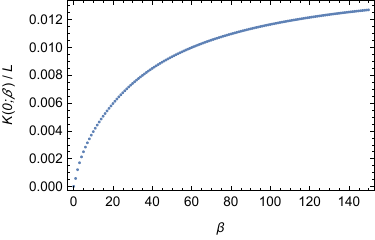}
    ~~\includegraphics[height=0.3\linewidth]{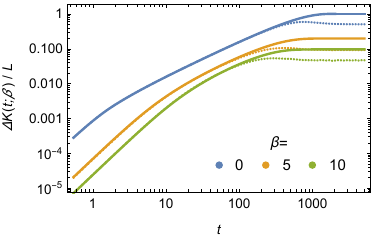}
    \caption{Krylov complexity as a function of inverse temperature  (left) and real time (right) for the GUE with $L=1024$ and $64$ realizations. The dots represent the numerical results and the solid curves represent analytical approximation from spectral complexity.}
    \label{fig:SFFKCGUETFDbeta}
\end{figure}

In the low temperature limit $1\ll\beta\ll L^{2/3}$, we may approximate the spectral density by the lower edge of the spectrum,
\begin{align}
	\rho(E)=\frac{L}{\pi}\sqrt{E},
\end{align}
where we have shifted the energy as $E\to E-2$ and considered $E\ll1$, {\it i.e.} the ``double scaled'' limit \cite{Cotler:2016fpe,Saad:2019lba}.
By evaluating \eqref{SFFEnergy}, we obtain the SFF
\begin{align}
	\avg{\abs{Z(\beta+it)}^2} = \frac{L}{4 \sqrt{2 \pi } \beta ^{3/2}}\text{erf}\left(\sqrt{\frac\beta 2} \frac{t}{L}\right) +\frac{L^2}{4 \pi  \left(\beta ^2+t^2\right)^{3/2}},
\end{align}
with $\text{erf}(z)=(2/\sqrt \pi)\int_0^z e^{-t^2} dt$ the error function.
Using the Ehrenfest theorem \eqref{Ehrenfest} and taking the double-time integral, we obtain a change of complexity of the form
\begin{align}
	C(t;\beta)
	=\left(\frac{L}{\beta }+\frac{t^2}{L}\right) \text{erf}\left(\sqrt{\frac\beta 2} \frac{t}{L}\right)
	+\sqrt{\frac{8}{\pi \beta}}\kd{ \frac t2 e^{-\frac{\beta  t^2}{2 L^2}}-t
	+\sqrt{\beta ^2+t^2}-\beta}
    -\frac{t^2}{L}.
\end{align}
Asymptotically, we find
\begin{align}\label{KCBeta}
	C(t;\beta)\to &
	\begin{cases}
		\sqrt{\frac{2}{\pi \beta ^3}}t^2,& t\ll\beta \\
		2\sqrt{\frac{2}{\pi \beta}}t,& \beta\ll t\ll L/\sqrt\beta \\
		\frac{L}{\beta }, & L/\sqrt\beta\ll t
	\end{cases}.
\end{align}
Similarly, the $\partial_t C(t;\beta)\to0$ when $t\to\infty$ and then $C(t;\beta)$ saturates at late times. 
We compare the spectral complexity $C(t;\beta)$ to the change of the numerical Krylov complexity $\Delta K(t;\beta)=K(t;\beta)-K(0;\beta)$ in the right panel of Fig.~\ref{fig:SFFKCGUETFDbeta}.
They have the same quadratic-to-linear growth behavior, but asymptote to different saturation values at late times. The initial value of Krylov complexity $K(0;\beta)$ also grows with the imaginary time $\beta$, as shown in the left panel of Fig.~\ref{fig:SFFKCGUETFDbeta}.

\section{Krylov complexity and chaos at late times}\label{sec:chaos}

Our calculations presented in the previous section demonstrate that the growth and saturation of Krylov state complexity are not significantly affected by the chaotic behavior of spectral statistics, such as spectral rigidity and level repulsion. 
Based on our analysis involving the Ehrenfest theorem, we find that the linear growth of Krylov complexity is mainly determined by the $n$ dependence of the Lanczos coefficients $b_n$ for small $n$, and the saturation is due to the discreteness of the spectrum.
In fact, in Sec.~\ref{sec:nonChaotic}, we demonstrate that the linear growth and saturation of the Krylov state complexity remain the same for non-chaotic spectra with uncorrelated levels.

The authors of Ref.~\cite{Balasubramanian:2022tpr} observed that the peak in Krylov complexity is sensitive to chaos. 
In this section, we focus on the late-time behavior of the transition probability $\abs{\phi_n(\beta+it)}^2$ and Krylov complexity $K(t;\beta)$. 
In the transition probability, we discover a universal rise-slope-ramp-plateau behavior that includes a long ramp related to chaos. 
We find that the long ramp in the transition probability is responsible for the peak in the Krylov complexity.

Our analysis of the late-time behavior of the transition probability and the Krylov complexity in the GUE is organized as follows. 
In Sec.~\ref{sec:NumericGUE}, we calculate the transition probability and Krylov complexity numerically and discuss their typical behavior. 
In Sec.~\ref{sec:firstlook}, we examine their fluctuations and give a first look at the ramp and the saturation.
In Sec.~\ref{sec:KCContinuumLimit}, we calculate the Krylov complexity in the continuum limit without taking the correlation between the spectral density and the polynomials of the Krylov basis into account. A part of this correlation is then included into the analysis in App.~\ref{sec:PolynomialEnsemble}.
In Sec.~\ref{sec:peak}, we examine the relationship between the ramp in the transition probability and the peak in the Krylov complexity.

\subsection{Numerical simulation}\label{sec:NumericGUE}

The Krylov complexity \eqref{KCTFD} in the ensemble average may be written as
\begin{align}
    \avg{K(t;\beta)}
    =\frac{D}{\avg{Z(2\beta)}}
    \sum_n n\avg{\abs{\phi_n(\beta+it)}^2}.
\end{align}
To study the chaotic dynamic from the Krylov space, we further examine the transition probability $\avg{\abs{\phi_n(\beta+it)}^2}$ for the ensemble average of the GUE. 
This quantity records the square of the amplitude of the wave function in the $n$-th Krylov basis. 
In addition, we examine its disconnected part $\abs{\avg{\phi_n(\beta+it)}}^2$ and connected part $\avg{\abs{\phi_n(\beta+it)}^2}_{\rm conn.}=\avg{\abs{\phi_n(\beta+it)}^2}-\abs{\avg{\phi_n(\beta+it)}}^2$. 
The former measures the square of the average amplitude of the wave function in the ensemble, and the latter measures the correlation between the fluctuation of the wave function over the ensemble. Other observables in the Krylov space, such as the quantity in \eqref{KC1}, can be split into their disconnected and connected parts in a similar manner. 
Our numerical simulation focuses on the case of $\beta=0$ as the peak of the Krylov complexity is already significant at this value.

\begin{figure}
    \centering
    \includegraphics[height=0.35\linewidth]{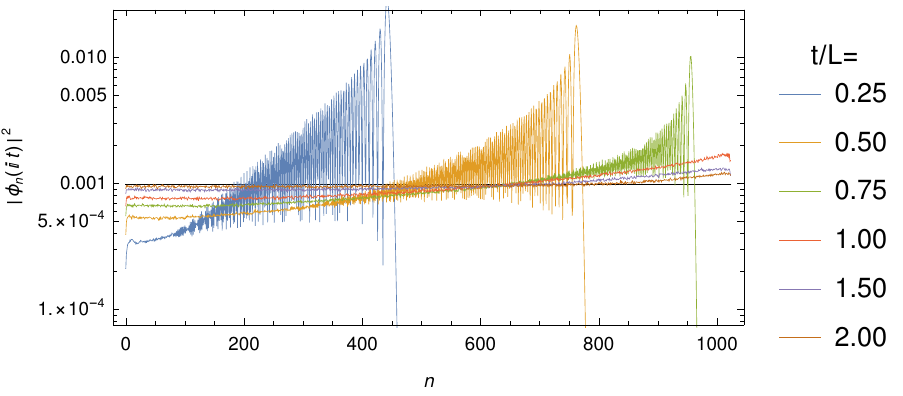}
    \caption{The snapshots of transition probability $\avg{|\phi_n(it)|^2}$ for different $t$ in chaotic system. We consider the GUE with $L=1024$ and $4096$ realizations. The black line denotes $1/L$.}
    \label{fig:SPsGUEWave}
\end{figure}

In Fig.~\ref{fig:SPsGUEWave}, we show numerical snapshots of the probability wave $\avg{\abs{\phi_n(it)}^2}$ at different times in the ensemble average of the GUE with a large $L$ and a huge number of realizations. 
We observe a shock wave with a long tail propagating forward and decaying during the evolution. 
Notably, it decays quickly when it reaches the end of the Krylov chain at $t\approx L$. 
The reflected wave is completely broken up. 
Once the shock wave passes by the Krylov chain, the global evolution of the probability wave is governed by diffusion. 
In particular, after $t\approx L$, the probability wave smoothly diffuses to a plateau value $1/L$, where the probability ramps up for $n\lesssim L/2$, and ramps down for $n\gtrsim L/2$.

\begin{figure}
    \centering
    \includegraphics[height=0.35\linewidth]{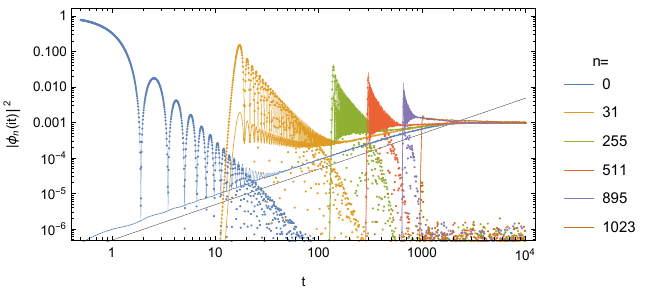}
    \includegraphics[height=0.35\linewidth]{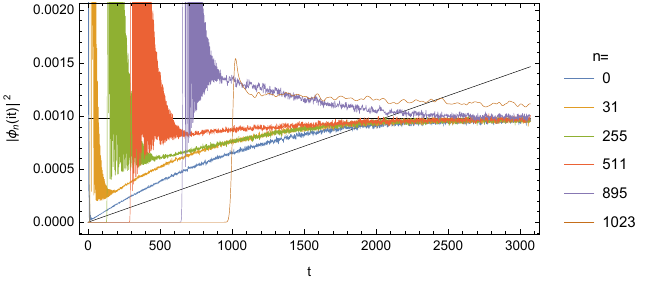}
    \caption{The time evolution of transition probability $|\phi_n(it)|^2$ for different $n$ in chaotic system. We consider the GUE with $L=1024$ and $4096$ realizations. The solid curves, the dashed curves, and the dots represent the full transition probability, the connected part, and the disconnected part respectively. The black lines denote $1/L$ and $t/(2L^2)$. The cloud of dots on the bottom is due the finite sampling.}
    \label{fig:SPsGUETFD}
\end{figure}

In Fig.~\ref{fig:SPsGUETFD}, we show the numerical time-evolution of the transition probability in the ensemble average $\avg{\abs{\phi_n(it)}^2}$, its disconnected part $\abs{\avg{\phi_n(it)}}^2$, and its connected part $\avg{\abs{\phi_n(it)}^2}_{\rm conn.}$ for various values of $n$. 
We observe a rise-slope-ramp-plateau behavior with a ramp-up for $n\lesssim L/2$ and a ramp-down for $n\gtrsim L/2$. 
This behavior has different features for different scales of $n$ compared to $L$. 
Specifically:
\begin{description}
\item [$n\ll L$] The rise-slope is short and is mainly contributed by the disconnected part. The ramp starts from a small step that depends on $n$, linearly increases, and then gradually slows down. The ramp is contributed by the connected part always, and gradually stops at a plateau time near $2L$.
\item [$n\lesssim L/2$] The rise-slope becomes longer and is partially contributed by the connected part. The ramp starts from an obvious step below the plateau, linearly increases with a smaller rate, and then gradually slows down. The plateau time is still near $2L$.
\item [$L/2\lesssim n<L$] The rise-slope is long and is mainly contributed by the connected part. The ramp starts from a step larger than the plateau, linearly decreases, and then gradually slows down. The plateau time is obviously greater than $2L$.
\end{description}

\begin{figure}
    \centering
    \includegraphics[height=0.35\linewidth]{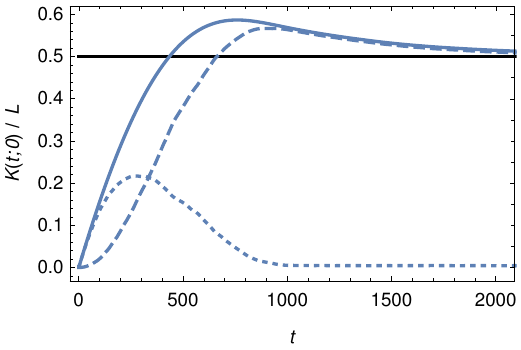}
    \caption{Krylov complexities as functions of time $t$ in chaotic system. We consider the GUE with $L=1024$ and $128$ realizations. The solid curves, the dashed curves, and the dots represent the full quantity, the connected part, and the disconnected part, respectively.}
    \label{fig:KCGUE}
\end{figure}

In Fig.~\ref{fig:KCGUE}, we show the numerical time-evolution of the Krylov complexity in the ensemble average $\avg{K(t;0)}$, as well as its disconnected part $\sum_n n \abs{\avg{\phi_n(it)}}^2$, and its connected part $\sum_n n \avg{\abs{\phi_n(it)}^2}_{\rm conn.}$. The peak is partially contributed by both the disconnected part and connected part.

\begin{figure}
    \centering
    \includegraphics[height=0.35\linewidth]{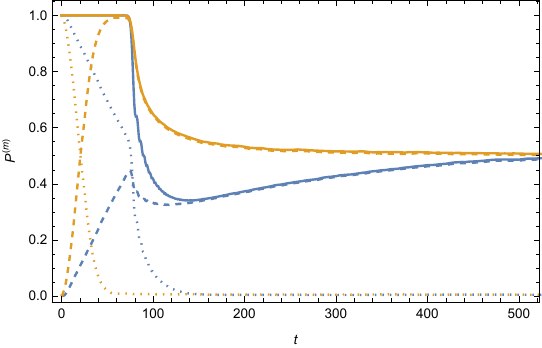}
    \caption{The $m$-site probability $P^{(m)}$ as a function of time $t$ in the chaotic systems (blue curves) and non-chaotic systems (orange curves) with $L=256$, $m=L/2$, $\beta=0$, and $128$ realizations. The solid curves, dashed curves, and dots represent the whole quantity, connected part, and disconnected part respectively.}
    \label{fig:mPGUETFD}
\end{figure}

A direct consequence of the ramp-up behavior in the transition probability is that nearly all observables in an $m$-dimensional Krylov subspace spanned by $\ke{\ket{O_n}}_{n=0}^{m-1}$ with $m\lesssim L/2$ will exhibit ramps after the dip time of $|\phi_m|^2$. For example, the $m$-site probability $P^{(m)}$ defined as
\begin{align}\label{SubObservable}
    P^{(m)}=\sum_{n=0}^{m-1} |\phi_n|^2,
\end{align}
ramps up after its slope in chaotic systems, as shown in Fig.~\ref{fig:mPGUETFD}.

In the following subsections, we will provide an analytical explanation for these numerical results from the perspective of chaos.

\subsection{A first look at late times}\label{sec:firstlook}

To explain the behavior of the transition probability, we first discuss the late-time plateau, and then the linear ramp appearing prior to the plateau.

Let us consider the behavior once  the plateau time is reached. Recall the plateau value of the transition probability and the saturation value of Krylov complexity in \eqref{AverageTFD}. 
To further estimate their fluctuations a long time after the plateau time $t_p$, we consider the long-time average of the probability correlation between sites $m$ and $n$ with time lag $dt$,
\begin{align}
    &(\abs{\phi_m(\beta+it)}^2 \abs{\phi_n(\beta+i(t+dt))}^2)_\infty-(\abs{\phi_m(\beta+it)}^2)_\infty (\abs{\phi_n(\beta+i(t+dt))}^2)_\infty \nn\\
    \approx & \sum_{pq}\abs{\avg{E_p|0}}^4\abs{\avg{E_q|0}}^4 e^{-2\beta (E_p+E_q)}e^{-idt(E_p-E_q)}\psi_m(E_p)\psi_n(E_p)\psi_m(E_q)\psi_n(E_q) \label{fluctuation}\\
    =& L^{-2}\abs{\bra{O_m}e^{-(2\beta+idt)H}\ket{O_n}}^2, \nn
\end{align}
where the subscript in $(\cdots)_\infty$ denotes the long-time average defined in \eqref{probabilityAverage}. For the first step in \eqref{fluctuation}, we use the property that $E_p-E_q=E_{p'}-E_{q'}$ holds only when $p=q,p'=q'$ or $p=p',q=q'$ in a chaotic spectrum \cite{Cotler:2017jue}. The final result in \eqref{fluctuation} is proportional to the transition probability between sites $m$ and $n$. 
From the characteristic curves \eqref{CharacteristicCurves}, we expect that it reaches a peak when $dt=\pm \abs{y(\epsilon m)-y(\epsilon n)}$, with $y(x)$ the coordinate in \eqref{WaveEq}. When $\beta=dt=0$, the correlation \eqref{fluctuation} is simply $L^{-2}\delta_{mn}$. Thus the local fluctuations take the form
\begin{align}
    [(\Delta\abs{\phi_n}^2)_\infty]^2=
    (\abs{\phi_n}^4)_\infty-[(\abs{\phi_n}^2)_\infty]^2=1/L^2 \, ,
\end{align}
which is of the same order as $[(\abs{\phi_n}^2)_\infty]^2$ in \eqref{AverageTFD}. From \eqref{fluctuation}, we may easily calculate the fluctuation of Krylov complexity with $\beta=0$ after saturation,
\begin{align}
    (\Delta K)_\infty^2=(K^2)_\infty-(K_\infty)^2
    =\sum_{m,n=0}^{L-1}mn L^{-2}\delta_{mn}
    =\frac{L}{3} + O(L^0) \, .
\end{align}
So the relative fluctuation is $(\Delta K)_\infty/K_\infty\approx 2/\sqrt{3L}$, which is negligible for large $L$. 
We conclude that  at $\beta=0$, the transition probability is not self-averaging after saturation, similarly to the SFF \cite{prange1997spectral}. 
On the other hand, we note that the Krylov complexity is indeed self-averaging after saturation, similar to the spectral complexity \cite{Iliesiu:2021ari}. 
So the relative fluctuation of the transition probability is stronger than the relative fluctuation of the Krylov complexity.

We move on to discussing the ramp before the plateau time. 
It is well known that the SFF, which is proportional to the survival probability, has a linear ramp. 
We expect that, due to the spectral rigidity, a linear ramp also appears in the transition probability $\abs{\phi_n(\beta+it)}^2$ for $n\neq0$ . 
For a first look at the linear ramp in the transition probability, we consider the two-point function \eqref{sine} in the GUE 
in the box approximation \cite{Liu:2018hlr,Cotler:2016fpe}
\begin{align}\label{sineBox}
    \avg{\rho(E_1)\rho(E_2)} 
    =\avg{\rho(E_1)}\avg{\rho(E_2)}
    +\avg{\rho(E)}\kd{\delta(s)-\frac{\sin^2\kc{Ls}}{\pi Ls^2}},
\end{align}
where we have sent $E_{1,2}\to E$ in the connected part as the sine kernel is localized at $s=0$. Moreover, we also sent $\avg{\rho(E)}\to \avg{\rho(0)}=L/\pi$ in the sine kernel, and have given an appropriate normalization to the sine kernel, such that the two-point function reduces to the one-point function $\avg{\rho(E_1)}$ if $E_2$ is integrated out.
From \eqref{WaveFunctionTFD}, the connected part of the probability on a given normalized polynomial $\psi(E)$ of low degree is
\begin{subequations}\label{RampFirstLook}
\begin{align}
    &~\frac1{L^2}\avg{\Tr\kd{\psi(H)e^{-(\beta+it)H}}\Tr\kd{\psi(H)e^{-(\beta-it)H}}}_{\rm conn.} \\
    =&~\frac1{L^2}\int_{-2}^2 dE\int_{-\infty}^{+\infty} ds \psi(E_1)\psi(E_2)\avg{\rho(E)}\kd{\delta(s)-\frac{\sin^2\kc{Ls}}{\pi Ls^2}}e^{-its-2\beta E} \label{RampFirstLook2} \\
    \approx &~\frac1{L^2}\int_{-2}^2 dE\int_{-\infty}^{+\infty} ds \psi(E)^2\avg{\rho(E)}\kd{\delta(s)-\frac{\sin^2\kc{Ls}}{\pi Ls^2}}e^{-its-2\beta E} \label{RampFirstLook3}\\
    =&~ \kc{\frac1{L}\int_{-2}^2dE \psi(E)^2\rho(E)e^{-2\beta E}}\min\ke{\frac{t}{2L^2},\frac1L} \label{RampFirstLook4}
\end{align}
\end{subequations}
In \eqref{RampFirstLook3}, we replace $\psi(E_1)\psi(E_2)\to\psi(E)^2$ by considering the that the sine kernel behaves as a narrow peak at $s=0$ with width $1/L$ and the polynomial $\psi(E)$ with low degree $n$ is relatively smooth compared with the sine kernel if $n\ll L$ and then the energy difference $s$ in $\psi(E_1)\psi(E_2)$ is negligible.
When $\beta=0$, the prefactor in \eqref{RampFirstLook4} is reduced to $1$ because of \eqref{Orthogonal}. So, at $\beta=0$, the connected part of the transition probability exhibits a ramp-to-plateau behavior, with the ramp $t/(2L^2)$, the plateau values $1/L$, and the plateau time $2L$, which approximately matches the numerical result for small $n$ in Fig.~\ref{fig:SPsGUETFD}. In App.~\ref{sec:GeneralRamp}, we show that the transition probability in a general Gaussian ensemble also exhibits a ramp. As we will show in Sec.~\ref{sec:nonChaotic}, the transition probability in a non-chaotic system have short ramp regions and different plateau times. So, the above ramp-to-plateau behavior characterizes the chaos in the Krylov space.

Note however that in \eqref{RampFirstLook}, we neglected two effects:
\begin{itemize}
    \item When $\psi(E)$ is a $n$-degree polynomial with $n\lesssim L$, the high-frequency oscillation prevent us from neglecting the energy different $s$ in $\psi(E_1)\psi(E_2)$.
    \item When the given polynomial $\psi(E)$ is taken as the polynomial $\psi_n(E)$ determined by the Lanczos coefficients, which are in principle determined by the spectrum $\ke{E_p}$ in a complicated way. The statistical correlation between the two polynomials and the two spectral density could also contribute to the connected part of the transition probability.
\end{itemize}

The first effect will leads to a $n$-dependent time shifting in the transition probability such that the ramp does not start from zero for finite $n$, as given in \eqref{RampPlateauContinuum} in App.~\ref{sec:ChaosBasis}.

The second effect is more complicated. In most of the literature, the statistical correlations between the two polynomials $\psi_n(E_1)\psi_n(E_2)$ and the density densities $\rho(E_1)\rho(E_2)$ in the ensemble average of the transition probability are neglected. Although we are unable to deal with all those complicated correlation in this paper, we still consider a most significant effect of these correlations step by step: 
In App.~\ref{sec:polynomials}, we will include an obvious effect of the correlation between the polynomial and spectral density, called the {\it confinement of polynomials}. In App.~\ref{sec:ApproximatePolynomial}, we will argue the approximated expressions of the reduced spectral density and polynomials after taking the effect of the confinement of polynomials into account.
In App.~\ref{sec:ChaosBasis}, we will find an improved expression of the transition probability with a ramp-to-plateau behavior in its connected part, as expected.

In the following subsections, we will first ignore the statistical correlation of the second effect and calculate the Krylov complexity in the first-order formalism of the continuum limit. The Krylov complexity already exhibits a peak as a result of the ramp in the transition probability at this level.

\subsection{Krylov complexity in the continuum limit}\label{sec:KCContinuumLimit}

\begin{figure}
    \centering
    \includegraphics[height=0.35\linewidth]{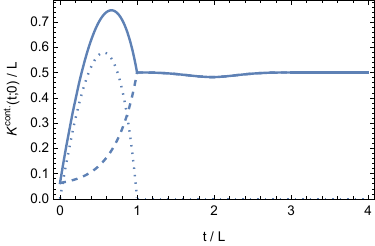}
    \caption{The Krylov complexity as a function of time in the continuum limit at $\beta=0$. The solid curves, the dashed curves, and the dotted curves represent the full quantity, the connected part, and the disconnected part, respectively.}
    \label{fig:KCCont}
\end{figure}

In this subsection, we derive the complexity in the first-order formalism of the continuum limit, while neglecting the statistical correlation between the polynomial $\psi_n(E)$ and the spectrum $\ke{E_p}$. According to \cite{Alishahiha:2022anw}, and using the solution \eqref{PolynomialC}, we may define a complexity operator on the energy basis in the continuum limit 
\footnote{We thank Souvik Banerjee for helpful discussions on this point.}.
However, in general, the polynomial in the continuum limit is not normalized to $1$, meaning that $(1/L)\int dE \abs{\Psi(E,x)}^2 \rho(E)=1/b(x)$ and $(1/\epsilon L)\int dx \abs{\Psi(E,x)}^2=\int dx /(\epsilon L b(x))$. Therefore, we should introduce a normalization factor $\sqrt{b(x)}$ to each $\Psi(E,x)$ before integrating over $x$. We obtain a complexity operator and total probability on the energy basis in the continuum limit,
\begin{align}
    J^{\rm cont.}(E_1-E_2)
    =&~\frac1{2\epsilon}\int_0^{\epsilon L} dx~x b(x)\kc{\Psi(E_1,x)^*\Psi(E_2,x)+\Psi(E_2,x)^*\Psi(E_1,x)}\nn\\
    =&~2\int_0^{y(\epsilon L)} dy~x(y)b(x(y)) \cos\kc{(E_1-E_2)y} ,\\
    P^{\rm cont.}(E_1-E_2)
    =&~\frac1{2}\int_0^{\epsilon L} dx~ b(x)\kc{\Psi(E_1,x)^*\Psi(E_2,x)+\Psi(E_2,x)^*\Psi(E_1,x)} \nn\\
    =&~2\epsilon\int_0^{y(\epsilon L)} dy~b(x(y)) \cos\kc{(E_1-E_2)y},
\end{align}
where we have taken the symmetric part under exchanging $E_1\leftrightarrow E_2$. 
Taking the limit value of the Lanczos coefficients \eqref{LanczosLimit}, we get 
\begin{align}\label{PolynomialContinuum1}
    \Psi(E,x)=\frac1{\sqrt{b(x)}}e^{iEy(x)}, \quad 
    b(x)=\sqrt{1-\frac{x}{\epsilon L}}, \quad 
    y(x)=L\kc{1-\sqrt{1-\frac{x}{\epsilon L}}}.
\end{align}
Then we get the complexity operator and the total probability in the continuum limit 
\begin{align}
    &J^{\rm cont.}(s)=\frac{\epsilon}{s^2}\kd{\left(L^2 s^2+6\right) \sinc^2\frac{L s}{2}-6}, \\
    &P^{\rm cont.}(s)=\epsilon L\sinc^2\frac{L s}{2},
    \quad s=E_1-E_2,
\end{align}
where we may set the overall factor as $\epsilon=1/L$ such that  $P^{\rm cont.}(0)=1$. 
The completeness relation \eqref{Complete} becomes $P^{\rm cont.}(s)$ in the continuum limit. 
The asymptotic behaviors of $J^{\rm cont.}(s)$ are
\begin{align}
    J^{\rm cont.}(s\to0)=\frac{L}{2}-\frac{L^3 s^2}{15}+O\left(s^3\right),\quad
    J^{\rm cont.}(s\to\infty)=-\frac{2 (\cos (L s)+2)}{L s^2},
\end{align}
where the $L/2$ will be the saturation value of the Krylov complexity at $\beta=0$, as a consequence of our normalization.
From \eqref{KCTFD}, the expectation value of the Krylov complexity in the continuum limit is 
\begin{align}
    &K^{\rm cont.}(t;\beta)=\frac{J^{\rm cont.}(t;\beta)}{P^{\rm cont.}(t;\beta)},
    \\
    &J^{\rm cont.}(t;\beta)=\frac1{L^2}\sum_{pq} J^{\rm cont.}(E_p-E_q)e^{it (E_p-E_q)-\beta(E_p+E_q)},\\
    &P^{\rm cont.}(t;\beta)=\frac1{L^2}\sum_{pq} P^{\rm cont.}(E_p-E_q)e^{it (E_p-E_q)-\beta(E_p+E_q)}.
\end{align}
With the two-point function in the GUE \eqref{sine}, we can calculate them in the ensemble average
\begin{align}
    \avg{J^{\rm cont.}(t;\beta)}=&~\frac1{L^2}\int_{-2}^{2} dE\int_{-\infty}^{+\infty}ds J^{\rm cont.}(s)\nn \\ 
    &~\ke{\rho_{\rm sc}(E)^2+\rho_{\rm sc}(E)\kd{\delta(s)-\frac{\sin^2(\pi\rho_{\rm sc}(E)s)}{\rho_{\rm sc}(E)(\pi s)^2}}} e^{its-2\beta E} ,\label{ComplexityContinuumLimit} \\
    \avg{P^{\rm cont.}(t;\beta)}=&~\frac1{L^2}\int_{-2}^{2} dE\int_{-\infty}^{+\infty}ds P^{\rm cont.}(s) \nn\\
    &~\ke{\rho_{\rm sc}(E)^2+\rho_{\rm sc}(E)\kd{\delta(s)-\frac{\sin^2(\pi\rho_{\rm sc}(E)s)}{\rho_{\rm sc}(E)(\pi s)^2}}} e^{its-2\beta E}.
\end{align}
where we have replaced the energies in the spectral density by their average $E$, since both $J^{\rm cont.}(s)$ and $P^{\rm cont.}(s)$ are localized around $s=0$ with a width $\sim20/L$. We will use the box approximation \eqref{sineBox} by replacing $\rho_{\rm sc}(E)\to \rho_{\rm sc}(0)$ in the sine kernel. The final result at $\beta=0$ is
\begin{align}
    \avg{K^{\rm cont.}(t;0)}=L\begin{cases}
        \frac{3 \pi  u^5-15 \pi  u^4+20 \pi  u^3+320 u^3-960 u^2+640 u+7 \pi }{-10 \pi  u^3+30 \pi  u^2-320 u+10 \pi +320}, 
        & u< 1\\
        \frac{3 u^5-15 u^4+20 u^3+15 u+7}{-10 u^3+30 u^2+30 u+10}, 
        & 1\leq u<2\\
        \frac{-3 u^5+45 u^4-260 u^3+720 u^2-945 u+519}{10 u^3-90 u^2+270 u-150}, 
        & 2\leq u<3\\
        \frac12, 
        & 3\leq u
    \end{cases}
    ,\quad u=\frac tL
\end{align}
whose configuration is plotted in Fig.~\ref{fig:KCCont}. $\avg{K^{\rm cont.}(t;0)}$ reaches a peak of value $0.75L$ at time $t\approx  0.7 L$ and nearly saturates to $0.5L$ when $t>L$. 
We further separate the contribution of the disconnected and connected parts of two-point function in the integrand of \eqref{ComplexityContinuumLimit} and show them in Fig.~\ref{fig:KCCont}. 
We see that the linear growth as well as the peak are mainly contributed by the disconnected part and the saturation is contributed by the connected part in this continuum limit. 
While the peak in the numeric Krylov complexity is partially contributed by the connected part, as shown in Fig.~\ref{fig:KCGUE}. 
The reason for discrepancy is that we neglect the fluctuation in the polynomial $\psi_n(E)$ coming from the fluctuation on levels. 
Those fluctuation transform some contribution from disconnected part to the connected part. 
We will come to this point in App.~\ref{sec:ChaosBasis}.

Furthermore, $\avg{K^{\rm cont.}(t;0)}$ grows as $\frac{7 \pi  }{10 (32+\pi )}L+\frac{16 (640+27 \pi ) }{5 (32+\pi )^2}t+O(t^2)$ initially, which are not reliable when compared to the early-time behavior in Sec.~\ref{sec:Earlytime}. 
They are affected by the loss of orthogonality, {\it e.g.} $\int dE\rho(E)\Psi(E,x)*1\neq0$ for $x\neq0$, which is an artifact of the continuum limit. 

\subsection{The peak of Krylov complexity}\label{sec:peak}

We can now explain the peak observed in the Krylov complexity, which appears slightly before the plateau time $t_p$ and much later than the dip time $t_d$ for small $n$. 
The transition probability for all small $n$ is in the ramp region, while the transition probability for all large $n$ is in the rise-slope region. 
Considering the weight $n$ in the Krylov complexity, the former is negligible while the latter dominates. 
Thus, the observed peak in the Krylov complexity is essentially the weighted combination of peaks in the rise-slope region of the transition probability at large $n$.

More precisely, we consider the saturated case where the shock wave reaches the $n$-th site at time $t$, and the transition probability before the $n$-th site is in its ramp region, i.e., $\avg{\abs{\phi_m(it)}^2}\gtrsim t/(2L^2)$ with $0\leq m < n$.
Due to probability conservation, we have $\avg{\abs{\phi_n(it)}^2}\lesssim 1-tn/(2L^2)$ at $\beta=0$, where the total probability is $1$. 
The shock wave travels according to \eqref{shock}. Then, the Krylov complexity is bounded above by
\begin{align}
    K\leq n(t)\kc{1-\frac{tn(t)}{2L^2}}+\frac{n(t)(n(t)-1)t}{4L^2}
    =\frac{t (2 L-t) \left(2 L^3-L t (2 t+1)+t^3\right)}{2 L^4},\ t\leq L , 
\end{align}
which has a peak at around $t=0.62L$.

We can also explain why the linear growth of Krylov complexity is not a result of chaos before its peak time. The dip times of the transition probabilities for different $n$ are mismatched, so the rise-slope of the transition probability in the shock wave always covers the ramps of other transition probabilities.

\section{Non-chaotic spectrum}\label{sec:nonChaotic}

\begin{figure}
    \centering
	\includegraphics[height=0.3\linewidth]{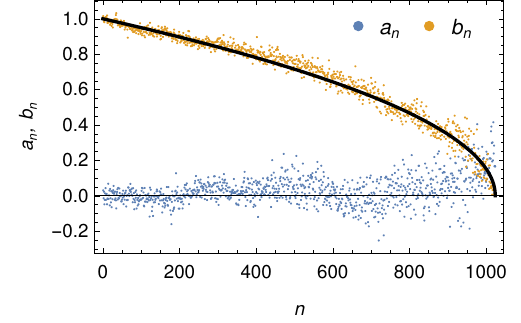}
	\includegraphics[height=0.3\linewidth]{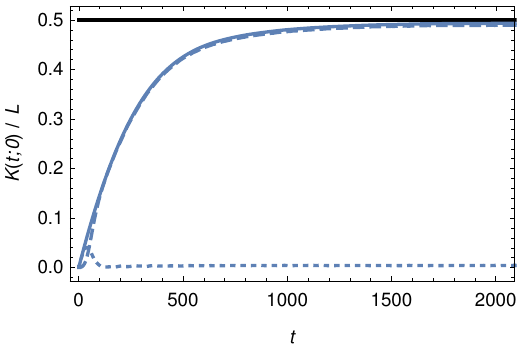}
	\includegraphics[height=0.35\linewidth]{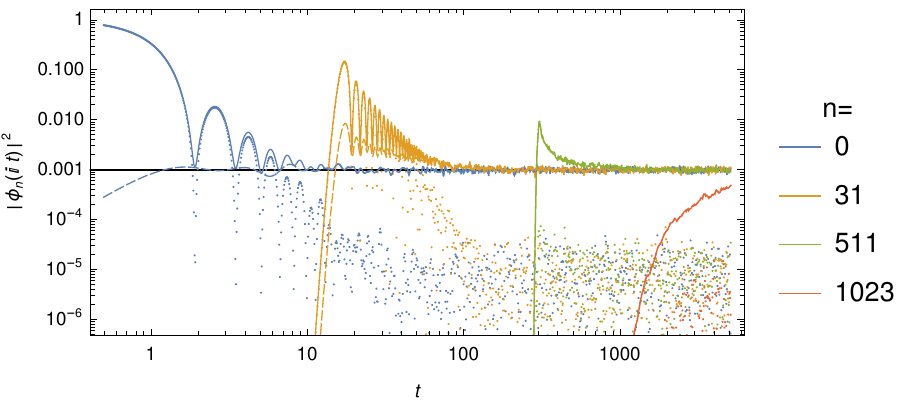}
    \caption{The Lanczos coefficients (upper-left), Krylov complexity (upper-right), and transition probability (lower) of the non-chaotic spectra with $L=1024$. In the last two plots, the solid curves, dashed curves, and dots represent the full quantity, connected part, and disconnected part, respectively, over $128$ realizations. The black line in the last plot denotes $1/L$.}
    \label{fig:NonChaotic}
\end{figure}

As a comparison, following the example given in \cite{Balasubramanian:2022dnj}, we can consider a uncorrelated spectrum whose density satisfies the same semicircle law \eqref{SemiCircle}. 
Technically, we randomly sample the energies individually from the spectra in the GUE with dimension $L$. 
We calculate the Lanczos coefficients, transition probability, and Krylov complexity numerically, as shown in Fig.~\ref{fig:NonChaotic}. The $m$-site probability is shown in Fig.~\ref{fig:mPGUETFD}.

The Lanczos coefficients approach the limit value \eqref{LanczosLimit} in ensemble average in consistence with the semi-circle law and \eqref{BlockDensity}. But they have strong fluctuation especially at large $n$. 

To study the behavior of transition probability, we consider the two-point function of uncorrelated spectrum
\begin{align}
    \avg{\rho(E_1)\rho(E_2)}=\rho(E)\delta(E_1-E_2)+\frac{L-1}{L}\avg{\rho(E_1)}\avg{\rho(E_2)},
\end{align}
where the spectral density in ensemble average obeys the semi-circle law  $\avg{\rho(E)}=\rho_{\rm sc}(E)$.
Similarly, considering the confinement and then decorrelating the polynomials and the spectral density, we obtain the SFF and the transition probability respectively
\begin{align}
	\avg{\abs{Z(\beta+it)}^2}
	=&~L+\frac{L(L-1)\abs{I_1(2\beta+2i t)}^2}{\beta^2+t^2},    \label{SFFNonChaotic}\\
    \avg{\abs{\phi_n(\beta+it)}^2}
    \approx&~\frac1{L^2}\int dE e^{-2\beta E}\avg{\psi_n(E)^2\rho^{(L-n)}(E)}+\frac{L-1}{L}\abs{\avg{\phi_n(\beta+it)}}^2\\
    = &~\frac1L+\frac{L-1}{L}\abs{\avg{\phi_n(\beta+it)}}^2
    \quad\text{when}\quad \beta=0.
\end{align}
The first terms will give a plateau and there is no ramp.
Since the Lanczos coefficients have strong fluctuations, the wave function $\phi_n(it)$, which is initially localized at $n=0$, quickly diffuses 
as it propagates, as observed in \cite{Balasubramanian:2022tpr} as well. 
Thus, when $n$ is not too large, the transition probability $\avg{\abs{\phi_n(it)}^2}$ exhibits rise-slope-plateau behaviors without any ramp, as shown in Fig.~\ref{fig:NonChaotic}. 
The plateau value is universally $1/L$, but the plateau time, {\it i.e.}, the time when its slope become smaller than $1/L$, highly depends on $n$ and is much earlier than $2L$. 
When $n$ is increasing and approaching $L$, the peak between the rise and the slope decays quickly and even ceases to exist. 
Finally, when the transition probability only gradually rises and approaches the plateau $1/L$ from below. 
The decay or even absence of the peak in the transition probability for $n\lesssim L$ is the result of probability conservation. 
Consider the time when $\avg{\abs{\phi_n(it)}^2}$ is rising, due to the absence of a ramp, all the $\avg{\abs{\phi_m(it)}^2}$ with $m\ll n$ have entered their plateau regions and all the $\avg{\abs{\phi_m(it)}^2}$ with $m\lesssim n$ have entered their slope region. 
In both cases, $\avg{\abs{\phi_m(it)}^2}\geq1/L$. From \eqref{ProbabilityConservation}, we get $\avg{\abs{\phi_n(it)}^2}<1-n/L$, which decreases with increasing $n$ and approaches its plateau value $1/L$ when $n\lesssim L$. 
Due to the absence of a ramp in the transition probability, we will not find any ramp in $m$-site observable \eqref{SubObservable}, as shown in Fig.~\ref{fig:mPGUETFD}. 
Following the strong fluctuations in the Lanczos coefficients, the polynomials $\psi_n(E_p;\ke{E})$ will get a stronger fluctuation at larger $n$ via the spectrum $\ke{E}$. 
So we observe that the connected parts of the transition probability $\avg{\abs{\phi_n}^2}_{\rm conn.}$ become dominated at large $n$.

Since the Lanczos coefficients are still around the limit value \eqref{LanczosLimit}. By plugging the SFF \eqref{SFFNonChaotic} into the Ehrenfest theorem \eqref{Ehrenfest} and doubly integrating over the time, we obtain the Krylov complexity at  the early time. In particular at $\beta=0$,
\begin{align} 
	K(t;0)\approx\, _1F_2\left(-\frac{1}{2};1,2;-4 t^2\right)-1
	\to\begin{cases}
		t^2, & t\ll 1\\
		\frac{16}{3 \pi }t, & 1\ll t\ll L
	\end{cases} \, ,
\end{align}
which agrees with the numerical result displayed in Fig.~\ref{fig:NonChaotic}. This linear growth has the same rate as the growth in RMT at early times. We thus confirm that, when the reference state is the maximally entangled state, the linear growth of Krylov state complexity does not characterize chaos in the spectrum \footnote{We distinguish  chaos in spectrum from chaos characterized by the exponentially decay of the OTOC, which is related to operators and locality \cite{Cotler:2017jue,Roberts:2016design}.}.

However, at late times, the Krylov complexity for a non-chaotic spectrum gradually slows down and reaches a plateau from below at $t\gtrsim L$, without going through a peak (see Fig.~\ref{fig:NonChaotic}), which is in contrast to the Krylov complexity in the GUE at late times (see Fig.~\ref{fig:KCGUE}). 
The absence of the peak in the Krylov complexity is due to the absence of the ramp in  the transition probability. Recall that $\avg{\abs{\phi_n(it)}^2}<1-n/L$ due to the early plateau time and probability conservation, which bounds the complexity contributions from probability at large $n$. To describe this in detail, let us consider  the saturated case, in which the shock reaches the $n$-th site at time $t$, {\it i.e.} $\avg{\abs{\phi_n(it)}^2}=1-n/L$ and $\avg{\abs{\phi_m(it)}^2}=1/L$ with $0\leq m< n$. With shock wave \eqref{shock}, the Krylov complexity is bounded from above by
\begin{align}
    K\leq n(t)\kc{1-\frac{n(t)}{L}}+\frac{n(t)(n(t)-1)}{2L}
    =\frac{L}{2} \left(1- \left(1-\frac{t}{L}\right)^4\right) + O(L^0),\quad t\leq L
    \end{align}
The bound monotonically grows from $0$ to $(L-1)/2$ when $n$ goes from $0$ to $L-1$ or $t$ goes from $0$ to $L$, and then it saturates. 
We may thus rule out the peak exceeding the saturated value $L/2$ in the Krylov complexity.

\section{The SYK model}\label{sec:SYK}

\begin{figure}
    \centering
	\includegraphics[height=0.3\linewidth]{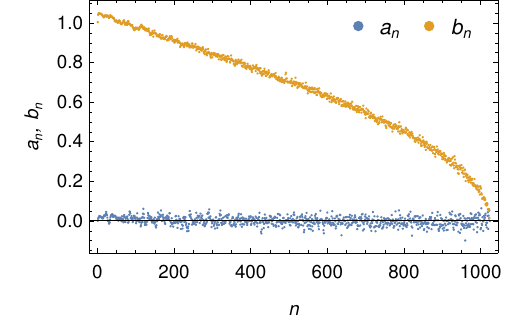}
	\includegraphics[height=0.3\linewidth]{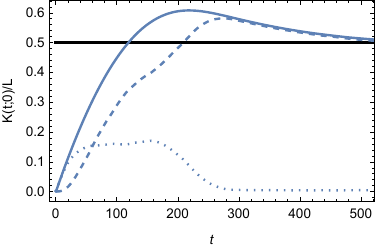}\\
	\includegraphics[height=0.35\linewidth]{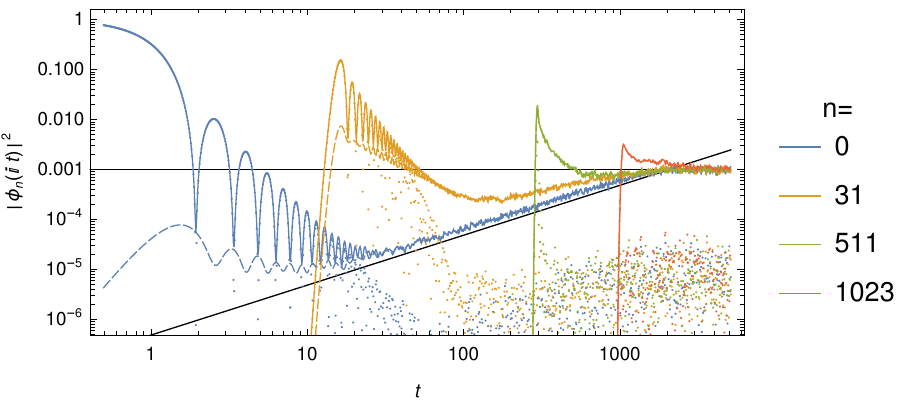}
    \caption{The Lanczos coefficients (upper-left), Krylov complexity (upper-right), and transition probability (lower) of the SYK$_4$ model with $N=22$, $L=1024$, and $128$ realizations. In the last two plots, the solid curves, dashed curves, and dots represent the whole quantities, connected parts, and disconnected parts, respectively. The black lines in the last plot denote $1/L$ and $t/(2L^2)$.}
    \label{fig:SYK}
\end{figure}

\begin{figure}
    \centering
	\includegraphics[height=0.3\linewidth]{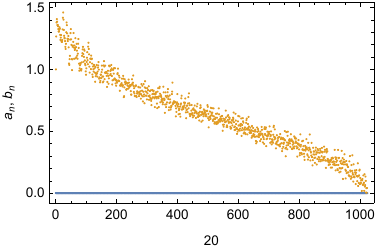}
	\includegraphics[height=0.3\linewidth]{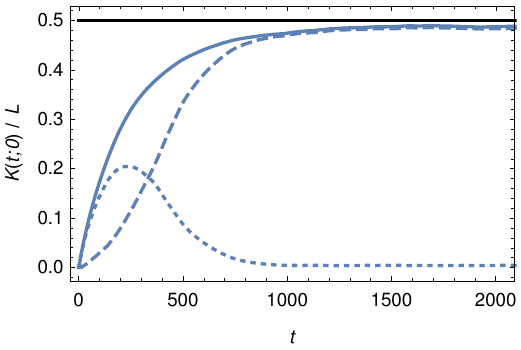}\\
	\includegraphics[height=0.35\linewidth]{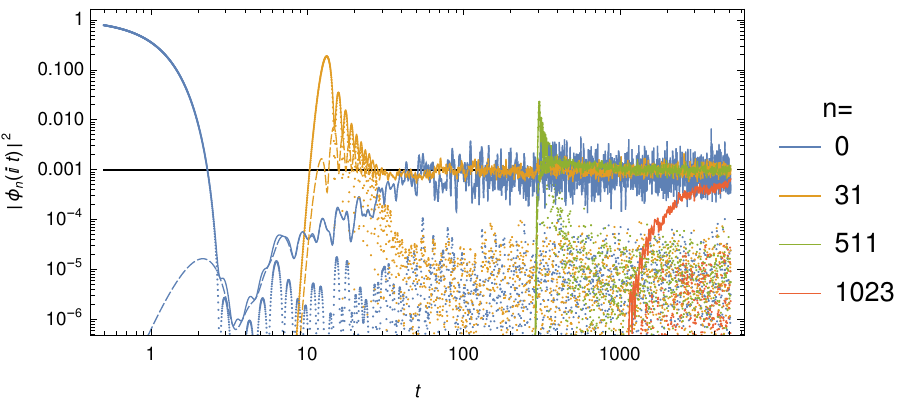}
    \caption{The Lanczos coefficients (upper-left), Krylov complexity (upper-right), and transition probability (lower) of the SYK$_2$ model with $N=20$, $L=1024$, and $128$ realizations. In the last two plots, the solid curves, dashed curves, and dots represent the whole quantities, connected parts, and disconnected parts, respectively. The black line in the last plot denotes $1/L$.}
    \label{fig:SYK2}
\end{figure}

Here we show that the universal rise-slope-ramp-plateau behavior also appears in the SYK model with quartic or higher fermion interactions. 
The SYK$_q$ model describes $N$ Majorana fermions with $q$-fermion random coupling, namely
\begin{align}
    H = i^{\frac q2} \sum_{1\leq j_1<j_2<\cdots<j_q\leq N} J_{j_1,...,j_q} \chi_{j_1}...\chi_{j_q}, \quad \avg{J_{j_1,...,j_q}^2} = \frac{2^{q-1}(q-1)!}{qN^{q-1}}\J^2,
\end{align}
and anti-commutation relation $\ke{\chi_{j},\chi_{k}}=\delta_{jk}$. 
We specify the value of $\J$ so that the first Lanczos coefficient is $b_1=1$. 

The SYK$_q$ model with $q\geq4$ has the same level spacing distribution as the RMT \cite{you2017sachdev,Garcia-Garcia:2016mno,Cotler:2016fpe}. The particle-hole symmetry determines the class of RMT statistics (GOE, GUE, and GSE) of each charge parity sector. 
For the GUE, we will consider $N \mod 8=2$ or $6$, where the spectrum consists of even and odd parity sectors which are mapped to each other by the particle-hole symmetry. 
Each sector corresponds to the GUE with dimension $L=2^{N/2-1}$. 

The density of state in the triple-scaled limit $N\gg q^2\gg1$ of the SYK model is \cite{Cotler:2016fpe}
$\avg{\rho(E)}\sim \J^{-1}e^{S_0}\sinh\kc{\pi\sqrt{2N(E-E_0)/(q^2\J)}}$
with $E_0$ the ground state energy and $S_0$ the zero temperature entropy.
Due to the scaled limit, only the lower edge of the spectrum is accessible. So a finite temperature is introduced to regularize the SFF, which is similar to Sec.~\ref{sec:LowTemperature}. 
The SFF of the SYK$_q$ model with the GUE statistic follows the integral \eqref{SFFEnergy} with the above spectral density. So the normalized SFF $\avg{\abs{Z(\beta+it)}^2}/\avg{Z(\beta)}^2$, {\it i.e.} the normalized survival probability,  exhibits a long ramp $\sim te^{-2S(\beta)}$ between the dip time $t_d\lesssim e^{S_0/2}$ and the plateau time $t_p\sim e^{S(\beta)}$, where $S(\beta)$ is the thermal entropy at the inverse temperature $\beta$  \cite{Cotler:2016fpe}. We would leave the analytical calculation of the transition probability and Krylov complexity in the future. Here we will numerically study them in the cases of $q=4,2$.

In the SYK$_4$ model, the spectral density is Gaussian law rather than semi-circle law \cite{Garcia-Garcia:2016mno,Maldacena:2016remarks}. 
We show the Lanczos coefficients, Krylov complexity, and transition probability at $\beta=0$ in Fig.~\ref{fig:SYK}.
The Lanczos coefficients $b_n$ decrease linearly for $n\ll L$ and decrease like $\sim\sqrt{1-n/L}$ for $n\lesssim L$. As \eqref{LanczosDifference} no longer holds, we cannot simply apply the Ehrenfest theorem as we did with \eqref{EhrenfestTFD}. Surprisingly, from the numerical Lanczos coefficients, we observe a modification of \eqref{LanczosDifference} as $b_{n+1}^2-b_n^2\approx C\delta_{n0}+An+B$ with $A\sim1/L^2$, $B\sim -1/L$ and $C\sim1$, which has an ``anomaly'' $C\delta_{n0}$ beyond the ``complexity algebra'' in \cite{Caputa:2021sib,Hornedal:2022pkc}. Plugging it into the Ehrenfest theorem \eqref{Ehrenfest}, we get 
\begin{align}\label{EhrenfestSYK}
    \partial_t^2K(t;\beta)\approx2\kc{\frac{C\abs{Z(\beta+it)}^2}{DZ(2\beta)}+AK(t;\beta)+B}.
\end{align}
For $\beta=0$, since $Z(it)$ decreases from $D$ and $K(t;\beta)\ll L$ in the early time, the first term dominates \eqref{EhrenfestSYK} and leads to a linear growth of order $t$ in the Krylov complexity. In Fig.~\ref{fig:SYK}, the transition probability exhibits the rise-slope-ramp-plateau behavior with ramp $\sim t/(2L^2)$, a common plateau time $t_p\approx 2L$ and a common plateau value $1/L$. The connected contributions to the rise-slope are more obvious. Due to the existence of a ramp in the transition probability, the Krylov complexity has a peak after its linear growth and before the plateau time.

The SYK$_2$ model is a many-body model of $N/2$ free Dirac fermions filling random matrix single-particle energy levels. 
The model is integrable. 
The many-body spectral density satisfies a Gaussian law. 
The levels have weak correlation and repulsion. The SFF has a short and exponential ramp compared to the dimension of Krylov space $L=2^{N/2}$, namely \cite{Winer:2020mdc}
\begin{align}
    \avg{\abs{Z(it)}}_{\rm conn.}^2\sim \exp\kc{\frac{\J\log N}\pi t},\quad 
    t<t_p\sim \frac N\J.
\end{align}
We show the Lanczos coefficients, Krylov complexity, and transition probability at $\beta=0$ in Fig.~\ref{fig:SYK2}
The Lanczos coefficients $b_n$ decrease faster than linearly for $n\ll L$ and $a_n=0$ due to the exact reversal symmetry of the spectrum. 
The transition probability exhibits the rise-slope-ramp-plateau behavior with a negligible ramp at finite $n$, a $n$-dependent plateau time $t_p$ and a common plateau value $1/L$. 
Due to the absent of a long ramp in the transition probability, the Krylov complexity does not have a peak. 

\section{Conclusion and outlook}\label{sec:conclusion}

We have studied the Krylov complexity in random matrix theory by mapping it to an effective Krylov spin chain model. Under the reasonable assumption of neglecting the statistical correlation between wavefunction and spectra, we find that the Krylov complexity satisfies an Ehrenfest theorem, in which the spectral form factor serves as a force to drive the complexity to grow. For random matrix theory, this also enables us to derive an analytical approximation of Krylov complexity, in terms of  spectral complexity \eqref{SpectralComplexity}. 
The analytical expression matches well with numerical calculations up to the linear growth region, while its asymptotic late-time value deviates from the actual saturation value. We attribute this deviation to the fact that neglecting the fluctuation of Lanczos coefficient is not valid at late times, as revealed by Fig.~\ref{fig:SPsGUETFD}. This deviation also shows the fundamental difference between the saturation of Krylov complexity and the saturation of spectral complexity at late times: the former relies on the discreteness of the spectrum \eqref{AverageTFD}, while the latter relies on level rigidity.

For early times, our generalized version of the Ehrenfest theorem is valid for the linear growth region of both chaotic and non-chaotic systems.  
This implies that the linear growth of complexity is not sufficient to discriminate chaotic and non-chaotic systems.

This further motivated us to study the quantum dynamics in the Krylov basis in general. We find that the transition amplitude, properly defined in a two-copy Hilbert space, shows a universal rise-slope-ramp-plateau behavior as  function of time. 
The linear ramp behavior is a robust indicator for chaotic systems, similarly to the spectral form factor. 
Therefore, unlike global observables that span over all of the Krylov basis, such as the Krylov complexity, any observable defined in the subspace of the Krylov space can reveal the linear ramp behavior that is unique for chaotic systems. Our study of the transition probability also explains the existence (absence) of the peak in the Krylov complexity in  chaotic (non-chaotic) systems discussed in \cite{Balasubramanian:2022tpr}.
Our analysis thus reveals a close relation between chaotic spectra and the Krylov state complexity. 

In this context, there remain some open questions for the future. 

First, the saturation value predicted by the Ehrenfest theorem is not accurate, and the error stems from the approximation that we neglect the fluctuations of the Lanczos coefficients in Eqs.~\eqref{LanczosLimit} and \eqref{EhrenfestTFD}. 
It will be interesting to improve the approximation by including more statistical correlations between the Lanczos coefficients and the Krylov wavefunction. 
A similar improvement can be obtained for the Krylov wavefunction $\phi_n$ at chain sites with large $n\sim L$. 

Second, in addition to the Ehrenfest theorem presented in this paper, further theorems in quantum mechanics may provide useful approximations or constraints for observables on the Krylov chain. For instance, the Robertson uncertainty relation provides a dispersion bound for Krylov complexity  \cite{Hornedal:2022pkc}. It will be interesting to investigate this bound in each region of Krylov state complexity discussed in this paper.

Third, for $n\gtrsim L/2$, the statistical correlation between the two polynomials in the transition probability makes the ramp-down behavior at late times difficult to analyze. In the numerical simulation, we observe that the late-time evolution of the transition probability is governed by diffusion rather than propagation. It would be more straightforward to describe the 
ramp-down behavior by writing down a diffusion equation for the Krylov chain.

Fourth, going beyond random matrix ensembles, it would be interesting to study Krylov state complexity in the SYK model \cite{Berkooz:2018jqr,Berkooz:2018qkz,Lin:2022rbf} and in holographic models such as JT gravity \cite{Saad:2019lba,Iliesiu:2021ari}, as well as in lattice models or spin models to benchmark the universal behavior studied in our paper.

Finally, let us highlight the fact that \eqref{KCTFD} provides an analytical expression for the Krylov complexity for the TFD state. This may serve as a starting point for constructing the gravity dual of Krylov complexity \cite{Chattopadhyay:2023fob}. As we show in Sec.~\ref{sec:chaos}, the connected part of the Krylov complexity plays an important role for its peak and saturation structure at time scales comparable to the dimension of the Hilbert space. This may be relevant in particular for describing the late-time evolution of black holes, where higher genus effects should be taken into account at finite $N$ in holography \cite{Iliesiu:2021ari}.

\section*{Acknowledgements}

We are grateful to Vijay Balasubramanian, Souvik Banerjee, Pawel Caputa, Adolfo del Campo, Pratik Nandy, and Dario Rosa for discussions. The research of J.~E.~and Z.-Y.~X.~is funded by DFG through the Collaborative Research Center SFB 1170 ToCoTronics, Project-ID 258499086—SFB 1170, as well as by Germany's Excellence Strategy through the W\"urzburg‐Dresden Cluster of Excellence on Complexity and Topology in Quantum Matter ‐ ct.qmat (EXC 2147, project‐id 390858490). Z.-Y.~X.~also acknowledges support from the National Natural Science Foundation of China under Grants No.~11875053 and No.~12075298.
S.-K.J~is supported by a startup fund at Tulane University.

\appendix

\section{Imaginary time evolution}\label{sec:Imaginary}

For imaginary time evolution, we consider the following continuum limit
\begin{align}
    x_n=\epsilon n,\quad \varphi(\tau,x_n)=(-1)^n\phi_n(\tau), \quad b(x_n-\epsilon/2)=b_n,\quad a(x)=a_n,
\end{align}
with a small spacing $\epsilon$ such that Eq.~\eqref{SchIm} becomes
\begin{align}\label{CSchIm}
\partial_\tau\varphi=&\epsilon^2b\varphi''+\epsilon^2b'\varphi'+(2b-a)\varphi + O(\epsilon^4).
\end{align}
In the continuum limit, the Ehrenfest theorem \eqref{Ehrenfest} becomes
\begin{align}\label{cEhrenfest}
    \partial_t^2J=\frac1{\epsilon^2}\partial_t^2\int |\varphi|^2 xdx
    =\int dx \kd{2 (b^2)' \abs{\varphi}^2-2 a'b\abs{\varphi}^2 + \epsilon^2bb''(\abs{\varphi}^2)'+\epsilon^2(b^2)'\abs{\varphi'}^2}
\end{align}
where the last two terms in the integrand may be neglected in the small $\epsilon$ limit.

To further write this into a typical Schr\"odinger equation in the continuous coordinates, we introduce a new wave function $\tilde\varphi(\tau,y)=b(x)^{1/4}\varphi(\tau,x)$ on coordinate $y=y(x)$ with $dy=dx/(\epsilon \sqrt{b})$,
and arrive at the following Schr\"odinger equation
\begin{align}\label{cSchEq}
	-\partial_\tau\tilde\varphi=-\partial_y^2\tilde\varphi+\tilde V\tilde\varphi,
\end{align}
with the potential
\begin{align}
	\tilde V(y)=\frac{B''(y)}{4 B(y)}-\frac{3 B'(y)^2}{16 B(y)^2}-2 B(y)+A(y)
	=\frac{\epsilon^2b''(x)}{4}-\frac{\epsilon^2b'(x)^2}{16 b(x)}-2 b(x)+a(x)
\end{align}
where $A(y)=a(x),\ B(y)=b(x)$, $A'(y) = \partial_y A(y), B'(y) = \partial_y B(y)$. 
The expectation value of operator $O$ transforms to \begin{align}
    \avg{O}
    =\sum_n O_n\abs{\tilde\varphi_n}^2
    =\frac1\epsilon\int_0^{\epsilon L} O(x) \abs{\tilde\varphi(x)}^2dx
    =\int_{y(0)}^{y(\epsilon L)} O(x(y))\abs{\tilde\varphi(y)}^2 dy.
\end{align}
We can read out another Ehrenfest theorem on the coordinate $y$ and real time $t=\Im\tau$,
\begin{align}
    \partial_t^2 \avg{y}= -\avg{\tilde V'(y)}.
\end{align}
However, the Krylov complexity $\avg{xL}$ is not directly related to $\avg{y}$.

\section{Spectral complexity for Gaussian ensembles}\label{sec:SpectralComplexityGaussian}

We may generalize the calculation of the spectral complexity in Sec.~\ref{sec:KCSFF} to the GOE and GSE. 
Instead of integrating the SFF over time, we directly use the spectral complexity \eqref{SpectralComplexity} and replace the two-point function \eqref{sine} with \cite{Liu:2018hlr}
\begin{align}\label{TwoPoints}
    \avg{\rho(E_1)\rho(E_2)}=\avg{\rho(E)}\delta(s)+\avg{\rho(E_1)}\avg{\rho(E_2)}-\frac12\Tr(k(E_1,E_2)k(E_2,E_1)),
\end{align}
where
\begin{align}
    & 
    k(E_1,E_2)=
    \begin{cases}
    \avg{\rho(E)}\begin{pmatrix}
        \sinc X& \sinc'(X) \\
        \Si(X) - \frac12\sgn(X) & \sinc X
    \end{pmatrix},& \text{GOE}\\
    \avg{\rho(E)}\begin{pmatrix}
        \sinc X& 0 \\
        0 & \sinc X
    \end{pmatrix},& \text{GUE}\\
    \avg{\rho(E)}\begin{pmatrix}
        \sinc(2X)& \sinc'(2X) \\
        \Si(2X) & \sinc(2X)
    \end{pmatrix},& \text{GSE}\\
    \end{cases}\\
    &E=(E_1+E_2)/2,\quad 
    X=\pi \avg{\rho(E)} s,\quad 
    s=E_1-E_2 ,\\
    &\sinc x=\frac{\sin x}{x},\quad 
    \Si(x)=\int_0^{x} \sinc x'dx',\quad 
\end{align}
and $\avg{\rho(E)}$ is taken as the semi-circle law \eqref{SemiCircle}.
Similarly to the case of JT gravity \cite{Iliesiu:2021ari}, the contact term in \eqref{TwoPoints} will not contribute to the complexity. In all of the three Gaussian ensembles, the spectral complexity is always written as
\begin{align}\label{SpectralIntegral}
    C(t;\beta)=\frac{2}{LZ(2\beta)}\int dE ds \avg{\rho(E_1)}\avg{\rho(E_2)} e^{-2\beta E} \kc{\frac{1-e^{its}}{s^2}}
    \kc{1-\tilde k\kc{\pi\avg{\rho(E)}s}},
\end{align}
with an ensemble-dependent kernel 
\begin{align}
    \tilde k(x)=
    \begin{cases}
    \sinc^2x+\frac1{x}(\Si(x)-\frac\pi2  \sgn x) (\sinc x-\cos x) ,& \text{GOE}\\
    \sinc ^2x,& \text{GUE}\\
    \sinc^2x+\frac1x\Si(x) (\sinc x-\cos x), & \text{GSE}
    \end{cases}.
\end{align}
The asymptotic behaviors of the kernel function $\tilde k(x)$ are
\begin{align}\label{KernelAsemptotic}
    \tilde k(x)\to 
    \begin{cases}
        1+O(x),& \abs{x}\ll1 \\
        f(x)\cos(2dx),& \abs{x}\gg1
    \end{cases} 
\end{align}
where $d=1,1,2$ for the GOE, GUE, and GSE respectively and $f(x)$ is a polynomial of finite degree. The integrand is therefore finite at $s=0$. For simplicity, we will focus on the $\beta=0$ case. 

We will split the spectral complexity into the disconnected part and the connected part, namely, $C(t)=C(t)_{\rm disc.}+C(t)_{\rm conn.}$. Obviously, the $1$ term and the $\tilde k(X)$ term in the integral factor $\kc{1-\tilde k\kc{X}}$ respectively contribute to the disconnected part $C(t)_{\rm disc.}$ and the connected part $C(t)_{\rm conn.}$.

The disconnected part is universal for those three ensembles, namely,
\begin{align}
    C(t)_{\rm disc.}=\, _1F_2\left(-\frac{1}{2};1,2;-4 t^2\right)-1
    \to\begin{cases}
		t^2, & t\ll 1\\
		\frac{16}{3 \pi }t, & t\gg 1
	\end{cases} \, ,
\end{align}
which exhibits a quadratic-to-linear growth in order $L^0$.

The connected part is highly dependent on the kernel. Since the kernel $\tilde k(\pi\avg{\rho(E)}s)$ is a narrow function of $s$ with width $1/\avg{\rho(E)}\sim 1/L$, we can replace $\avg{\rho(E_1)}\avg{\rho(E_2)}\to \avg{\rho(E)}^2$ and extend the domain of integral w.r.t. $s$ to the real axes. By integrating out $s$, we get the connected part
\begin{align}
    C(t)_{\rm conn.}=\int dE~\frac{4\pi^2\avg{\rho(E)}^3}{3L^2} g\kc{\frac{t}{2\pi \avg{\rho(E)}}}
\end{align}
where
\begin{align}
&g(u)_{\rm GOE}
=\begin{cases}
\frac{1}{12} \left(\left(-12 u^3+9 u+3\right) \log (2 u+1)+2 u (u (17 u-24)-3)\right), & u\leq 1 \\
\left(u^3-\frac34u+\frac{1}{4}\right) \log (2 u-1)+\left(- u^3+\frac34 u+\frac14\right) \log (2 u+1)& ~~~\\ 
\hspace{0.5\linewidth}
+(u-3) u+\frac{1}{3} , & u>1 \\
\end{cases},\\
&g(u)_{\rm GUE}
=\begin{cases}
u^2(u-3), & u\leq 1 \\
1 - 3u, & u> 1 \\
\end{cases},\\
&g(u)_{\rm GSE}
=\begin{cases}
\frac{1}{24} \left(u (u (17 u-60)-12)-6 \left(u^3-3 u+2\right) \log \abs{1-u}\right), & u\leq 2 \\
\frac{2}{3}-3 u, & u>2 \\
\end{cases}
\end{align}
Only for the GUE, we can perform the integral w.r.t. $E$ analytically and reproduce \eqref{KGUETFD} in the main text. For other ensembles, we have to compute it numerically. The results of spectral complexity are shown in Fig.~\ref{fig:KCEns}. 

At the early time $t\ll L$, as $g(u)=-3u^2+O(u^3)$, the connected part scales as $C(t)_{\rm conn.}\sim t^2/L$ and is negligible compared to $C(t)_{\rm disc.}$, as discussed in Sec.~\ref{sec:KCSFF}.

At the late time $t\gtrsim L$, the growth of the spectral complexity will slow down, stop, or even rebound in different ensembles, due to the cancellation between the disconnected part and connected part, as explained in \cite{Iliesiu:2021ari} and the following text. Let's first assume that $\tilde k(x)$ is holomorphic function of $x$. Based on the asymptotic behavior \eqref{KernelAsemptotic}, when $t>2 d L\geq 2\pi d\avg{\rho(E)}$, the integrand $e^{its}(1-\tilde k(\pi\avg{\rho(E)}s))$ is analytic on the upper half-plane and the real axis of $s$. So we can close the contour of integral $\int ds$ by using the arc at complex infinity in the upper-half plane, such that the integral of those terms vanishes and the spectral complexity saturates to a time-independent value given by the other terms. 
However, for the GOE, the $\sgn x$ in $\tilde k(x)$ is not analytic. 
As a result, the $C(t)$ for the GUE approaches $\log t$ instead of a constant.

\section{Krylov approach in ensemble average} \label{sec:PolynomialEnsemble} 

\subsection{Formal Krylov approach in ensemble average}

Here we give the formal expressions of the transition probability and the Krylov complexity in the ensemble average of Gaussian ensembles. As explained in the main text, to construct the Krylov space of the maximally entangled state $\ket0$, we can consider non-degenerate spectrum $\ke{E_p}_{0\leq p\leq L-1}$. 
The Krylov space is spanned by the states $\L^n\ket0$ for $0\leq n\leq L-1$. 
The components of the $n$-th state on the energy basis of $\L$ are the $n$-th rows of the Vandermonde matrix
\begin{align}
    \mathbf V=\ke{E_p^n}_{0\leq n,p\leq L-1}
\end{align}
times $1/\sqrt L$. 
So the Krylov basis $\ket{O_n}$ is obtained from the orthogonalization of $\mathbf V$, namely
\begin{align}
    \mathbf \Psi=\ke{L^{-1/2}\psi_n(E_p)}_{0\leq n,p\leq L-1}= \mathbf{ CV,\quad \Psi \Psi}^T=\mathbf \Psi^T\mathbf \Psi=\mathbf I,
\end{align}
where $\mathbf C$ is a lower-triangular matrix and $\mathbf I$ is the identity matrix. From the moments and Hankel matrix
\begin{align}
    \mu_j=\sum_p E^j_p,\quad    \mathbf A^{(m,n)}=\ke{\mu_{i+j}}_{0\leq i\leq m-1,0\leq j\leq n-1},
\end{align}
one can also derive the orthogonal polynomials by
\begin{align}
    \psi_n(E)=\frac1{\sqrt{h_n}\det \mathbf A^{(n,n)}}\det
    \begin{pmatrix}
    \mathbf A^{(n,n+1)}\\
    1 ~ E ~ E^2 ~ \cdots ~ E^n
    \end{pmatrix}.
\end{align}
If the spectrum is symmetric, one can also divide the Lanczos coefficients from the Hankel determinant
\begin{align}
    D_n=\det \mathbf A^{(n,n)},\quad D_0=1,\quad 
    b_n^2=\frac{D_{n-1}D_{n+1}}{D_n^2}
\end{align}
The transition probability and Krylov complexity are respectively
\begin{align}
    \abs{\phi_n}^2 =  \frac{\kc{\mathbf{\Psi F \Psi}^T}_{nn}}{\Tr\mathbf F},\quad 
    K=\frac{\Tr[\mathbf {K \Psi F \Psi}^T]}{\Tr\mathbf F},
\end{align}
where the evolution matrix and complexity matrix are respectively
\begin{align}
    \mathbf F= \ke{e^{-\tau E_p-\tau^* E_q}}_{0\leq p,q\leq L-1},\quad 
    \mathbf K=\ke{n\delta_{mn}}_{0\leq m,n\leq L-1}.
\end{align}

In an ensemble average, they become
\begin{align}\label{KCEnsemble}
    \avg{\abs{\phi_n}^2} =  \int DE\frac{\kc{\mathbf{\Psi F \Psi}^T}_{nn}}{\Tr\mathbf F},\quad 
    \avg{K}=\int DE\frac{\Tr[\mathbf{K \Psi F \Psi}^T]}{\Tr\mathbf F},
\end{align}
For $\tilde\beta$-ensemble, the measure of the spectrum is
\begin{align}\label{MeasureBeta}
    \int DE=\frac1 {Z_{\tilde \beta}}\int dE_0dE_1\cdots dE_{L-1} \exp\kc{-\frac{\tilde\beta L}{4}\sum_{0\leq p\leq L-1} E_p^2}\prod_{0\leq p<q\leq L-1}\abs{E_p-E_q}^{\tilde\beta} 
\end{align}
which is normalized to $1$ and we have rescaled the potential such that the domain of spectrum is $[-2,2]$ in expectation. We will work at the GUE $(\tilde\beta=2)$ for simplicity. It can be solved with the oscillator wave functions
\begin{align}
    \varphi_n(x)=\frac{1}{\sqrt{2^n n!\sqrt\pi}}\exp \left(-\frac{x^2}{2}\right) H_n(x)
\end{align}
where $H_n(x)$ is the Hermite polynomial. By using the kernel
\begin{align}
    K_n(x,y)=\sum_{m=0}^{n-1}\varphi_m(x)\varphi_m(y)
\end{align}
one can write down the one point function and two point function 
\begin{align}
    \avg{\rho(E)}=K_L(E,E),\quad
    \avg{\rho(E_1,E_2)}=K_L(E_1, E_2)^2.
\end{align}
In the large $L$ limit, they become  the semi-circle law and sine kernel
\begin{align}
    \avg{\rho(E)}=\frac{L}{2\pi}\sqrt{4-E^2},\quad
    \avg{\rho(E_1,E_2)}=\sinc(\pi L(E_1-E_2)).
\end{align}

Since $\psi_n(E_p)$ is a complicated function of the spectrum of degree $n$, the ensemble average of probability $\avg{\abs{\phi_n}^2}$ is a complicated summation of maximally $2n$-point functions and thus difficult to solve analytically. 
So we will approximate $\psi_n(E_p)$ by the polynomials $\psi_n(E)$ given by the limit value \eqref{LanczosLimit} and take its confinement into account.

\subsection{General linear ramp from Gaussian ensemble}\label{sec:GeneralRamp}

Similar to the SFF, the ramp in the transition probability does not rely on the specific sine kernel. 
Here, we work at $\beta=0$ for simplicity. 
Since the $\psi_n(E)$ in $\abs{\phi_n}^2$ is a $n$-degree polynomial of $E$, we can consider the generating function in the Gaussian ensemble average $\avg{\cdots}$, namely
\begin{align}
    \avg{\Tr[e^{-it_1 H}]\Tr[e^{it_2 H}]}
    =\int D\rho \int dE_1dE_2 \rho(E_1)\rho(E_2)e^{-it_1E_1+it_2 E_2}
    e^{-S[\rho]},
\end{align}
where $S[\rho]$ is an effective action of spectral density $\rho$ from Gaussian ensemble. 
The saddle point of $S$ is the semi-circle $\rho_s$. 
Considering fluctuation $\rho=\rho_s+\delta \rho$ and expanding $S$ around its saddle, one finds the quadratic term \cite{Cotler:2017jue}
\begin{align}
    \delta S
    =-L^2\int^\Lambda dE_1dE_2 \delta\rho(E_1)\delta\rho(E_2)\log\abs{E_1-E_2} 
    =\frac {L^2}{\Lambda^2}\sum_{n} \delta\rho_n\delta\rho_{-n}\frac{\text{Si}(n\pi)}{n\pi }, 
\end{align}
where $\delta\rho_n=\int^\Lambda dE \delta\rho(E) e^{iEs_n}$ is the Fourier transformation from a finite domain of spectrum $E\in[-\Lambda/2,\Lambda/2]$ to discrete $s_n=2\pi n/\Lambda$ with $n\in\mathbb N$, $\text{Si}(x)=\int_0^x (\sinc t) dt$, and $\sinc(x)=(\sin x)/x$. 
Then the connected part of the generating function is related to the two-point function of $\delta\rho$, namely
\begin{align}
    &\int D\delta\rho \sum_{mn}\delta\rho_m\delta\rho_n \sinc(\pi m+\Lambda t_1/2)\sinc(\pi n-\Lambda t_2/2) e^{-\delta S}\\
    =&\frac{L^2}{\Lambda^2}\sum_n\frac{n\pi}{\text{Si}(n\pi)}\sinc(-n\pi+\Lambda t_1/2)\sinc(n\pi-\Lambda t_2/2) \\
    \approx& \frac{L^2\sqrt{t_1t_2}}{\pi \Lambda}\sinc(\Lambda(t_1-t_2)/2),\quad \text{when}\quad \abs{t_1-t_2}\ll 1/\Lambda.
\end{align}
By replacing $t_1\to t+t_1$ and $t_2\to t-t_2$ and letting $\abs{t_{1,2}}\ll t$, we obtain a linear ramp on $t$ in the connected part of the generating function
\begin{align}\label{GeneratingRamp}
    \avg{\Tr[e^{-i(t+t_1) H}]\Tr[e^{i(t-t_2) H}]}_{\rm conn.}
    \approx\frac{L^2t}{\pi \Lambda}\sinc(\Lambda(t_1+t_2)/2).
\end{align}
Thus, by taking the derivative of \eqref{GeneratingRamp} with respect to $t_{1,2}$, we will obtain a linear ramp.

\subsection{Confinement of polynomials}\label{sec:polynomials}

\begin{figure}
    \centering
    \includegraphics[width=\linewidth]{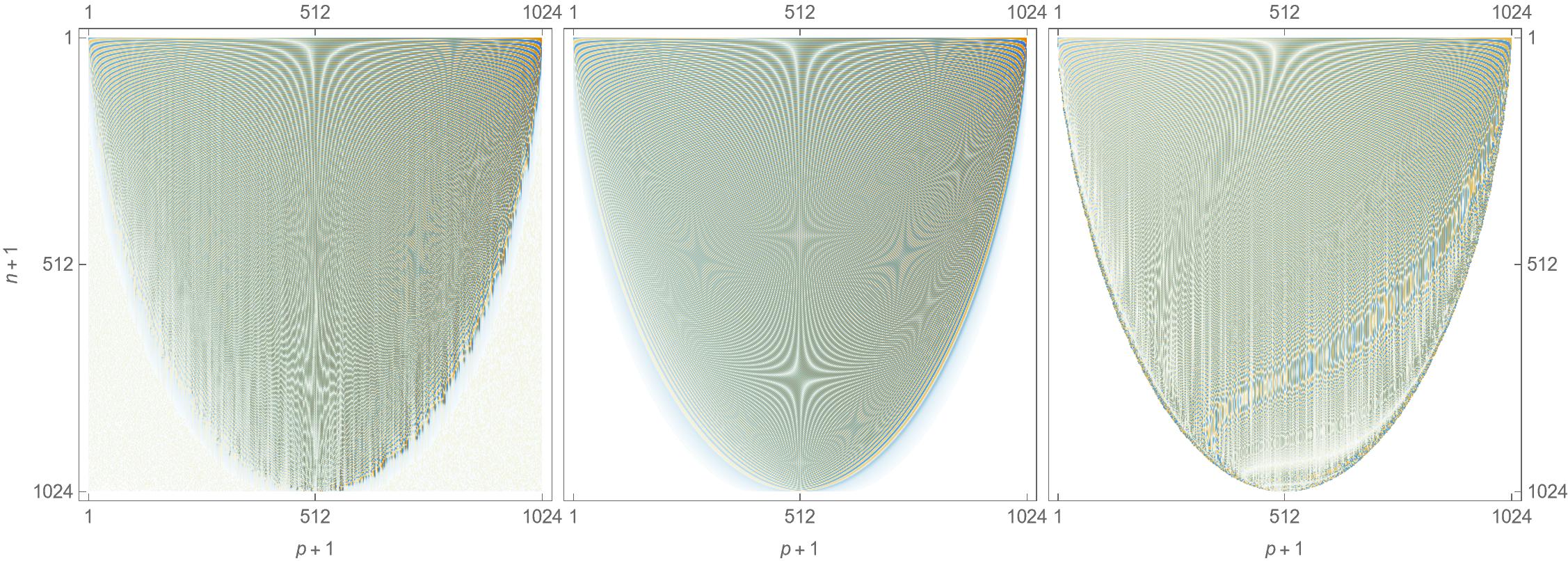}
    \caption{Polynomials $\psi_n(E_p)$ as functions of $n$ and $p$, where $L=1024$. Left: the polynomials in the Krylov space of TFD state in one realization of the GUE. Middle: the polynomials from \eqref{LanczosLimit}. Right: the deformed Chebyshev polynomials \eqref{PolynomialChebyshev} on a GUE spectrum  within $-2\sqrt{1-n/L}<E_p<2\sqrt{1-n/L}$.}
    \label{fig:Polynomials}
\end{figure}

The transition probability in ensemble average is defined as
\begin{align}\label{TPEnsemble}
    &~\abs{\phi_n(\beta+it)}^2=\frac1{L^2}\avg{\sum_{pq}\psi_n(E_p;\ke{E})\psi_n(E_q;\ke{E})e^{-\beta(E_p+E_q)-it(E_p-E_q)}} \\
    =&~\frac1{L^2}\avg{\int dEdE' \psi_n(E;\ke{E})\psi_n(E';\ke{E}) e^{-\beta(E+E')-it(E-E')} \rho(E)\rho(E')}
\end{align}
where $\psi_n(E_p;\ke{E})$ is a polynomial of its first argument $E_p$ and the polynomial is determined by the orthogonality relation \eqref{Orthogonal}, or equivalently, the Lanczos algorithm \eqref{Algorithm}\eqref{polynomial}. 
Since the orthogonalization is determined by the whole spectrum $\ke{E}=\ke{E_0,E_1,\cdots,E_{L-1}}$, where we have shown the dependence on $\ke{E}$ explicitly, the ensemble average of the transition probability \eqref{TPEnsemble} essentially relies on the multi-point (more than two points) function of the spectral density.

In principle, the orthogonalization before the ensemble average is different from the orthogonalization after the ensemble average. In other words, the former case corresponds to applying the Lanczos algorithm \eqref{Algorithm} for each realization of ensemble and taking their average finally, which is what we are simulating in Sec.~\ref{sec:NumericGUE}. The latter case corresponds to a single Lanczos algorithm \eqref{Algorithm} for the ensemble average of the inner product. 

In this paper, we are not able to solve \eqref{TPEnsemble} exactly. Instead, in this subsection, we try to develop an approximate method by including the a significant effect in the level correlation between the argument $E_p$ and the spectrum $\ke{E}$ in the polynomial $\psi_n(E_p;\ke{E})$, called the confinement of polynomials. 

Instead of thinking $\psi_n(E';\ke{E})$ as a function of a continuous energy $E'$, we only have to consider the its dependence on the discrete levels $E'=E_p\in \ke{E}$ with $p=0,1,\cdots,L-1$. 
We show $\psi_n(E_p;\ke{E})$ as a function of $E_p$ and $n$ for a realization of the GUE in the left panel of Fig.~\ref{fig:Polynomials}. 
An important feature is that the $\psi_n(E_p;\ke{E})$ almost vanishes for $|E_p|\gtrapprox 2\sqrt{1-n/L}$, which is different from the profile of $\psi_n(E';\ke{E})$ in the continuous domain of spectrum, namely $[-2,2]$ in our convention. 
In other words, $\psi_n(E';\ke{E})$ passes its zeros on those $E'=E_p \not\in [-2\sqrt{1-n/L},2\sqrt{1-n/L}]$. 
We say that $\psi_n(E_p;\ke{E})$ is confined in $[-2\sqrt{1-n/L},2\sqrt{1-n/L}]$. 

The confinement of $\psi_n(E_p;\ke{E})$ is due to the decrease of $b_n$ as a sequence of $n$, as explained as follows. 
Let us send $b_n$ of the corresponding $n$ in $\mathbf{L}$ to $0$ and denote the resulting matrix as $\mathbf{\tilde L}$. 
The spectra of $\mathbf L$ and $\mathbf{\tilde L}$ are close. 
Since $\mathbf{\tilde L}=\mathbf{L}^{(n)}\oplus \mathbf{R}^{(L-n)}$, where $\mathbf L^{(n)}$ and $\mathbf{R}^{(L-n)}$ are the upper-left block and the lower-right block respectively, the whole spectrum is the union of the spectra of $\mathbf{L}^{(n)}$ and $\mathbf{R}^{(L-n)}$.  
So, we can separate the summation over the spectrum into two parts
\begin{align}
    \sum_p = \sum_p^{(n)}{}' + \sum_p^{(L-n)},
\end{align}
where $\sum_p^{(n)}{}'$ sum over the levels from $\mathbf L^{(n)}$ and $\sum_p^{(L-n)}$  sum over the levels from $\mathbf R^{(L-n)}$. Correspondingly, the spectral density \eqref{LanczosToDensity} could be approximately divided into two parts
\begin{align}\label{DensityDivide}
    \rho(E)=\bar\rho^{(n)}(E)+\rho^{(L-n)}(E),
\end{align}
where the spectral density $\bar\rho^{(n)}(E)$ of $\mathbf{L}^{(n)}$ and the spectral density $\rho^{(L-n)}(E)$ of $\mathbf{R}^{(L-n)}$ are given by the integrals \eqref{LanczosToDensity} with modified bounds $\int_0^{n/L} dx$ and $\int_{n/L}^1 dx$, which are normalized to $n$ and $L-n$, respectively.

Recall that $p_n(E';\ke{E})=\det(E'-\mathbf{L}^{(n)})$ from \eqref{polynomial}. 
So $p_n(E_p;\ke{E})$ automatically vanishes on the spectrum of $\mathbf{L}^{(n)}$. Then we can reduce the summation 
\begin{align}\label{ReducedSum}
    \sum_p p_n(E_p;\ke{E})^{j} f(E_p)
    \approx\sum_p^{(L-n)} p_n(E_p;\ke{E})^j f(E_p),
\end{align}
where the power is taken as $j=1,2$, the $\sum_p^{(L-n)}$ is taken on the spectrum of $\mathbf{R}^{(L-n)}$, and $f(E_p)$ is an arbitrary function of $E_p$.

After the confinement of polynomials is taken into account, the reduced summation \eqref{ReducedSum} in the ensemble average given in \eqref{MeasureBeta} is
\begin{align}
    \int DE\sum_p p_n(E_p;\ke{E})^j f(E_p)
    \approx&\int DE\sum_p^{(L-n)} p_n(E_p;\ke{E})^j f(E_p) \nn \\
    =& \int DE\int dE'p_n(E';\ke{E})^jf(E')\sum_p^{(L-n)} \delta(E'-E_p)\nn\\
    \approx&\int dE' \avg{p_n(E')}^j f(E')\avg{\rho^{(L-n)}(E')}
\end{align}
At the last step, we consider the average separately by neglecting the correlation between each term and define $\avg{\rho^{(L-n)}(E)}$ and $\avg{p_n(E)}$ as the reduced spectral density and the polynomials in ensemble average respectively. The ensemble average of the reduced one-point function is
\begin{align}\label{Sub1pt}
    \avg{\rho^{(L-n)}(E')}=\avg{\sum^{(L-n)}_p\delta(E'-E_p)}.
\end{align}
The ensemble average of the normalized polynomial $\avg{\psi_n(E')}$ is proportional to $\avg{p_n(E')}$. But the normalization is slightly subtle. Due to the trick of setting $b_n=0$ and replacing $\mathbf L$ by $\tilde {\mathbf L}$, the original norm $\prod_{m=1}^n b_m^2$ vanishes for each realization. We have to renormalize the polynomial in the ensemble average
\begin{align}\label{FixNorm}
    \frac1L\int_{-2\sqrt{1-n/L}}^{2\sqrt{1-n/L}} dE\avg{\psi_n(E)}^2\avg{\rho^{(L-n)}(E)}=1.
\end{align}
with the reduced spectral density $\avg{\rho^{(L-n)}(E)}$.
By considering the confinement and the renormalization of polynomials, we take the ensemble average as 
\begin{align}\label{PolynomialEnsemble}
    \int DE\sum_p \psi_n(E_p;\ke{E})^j f(E_p)
    \approx
    \int_{-2\sqrt{1-n/L}}^{2\sqrt{1-n/L}} dE\avg{\psi_n(E)} \avg{\rho^{(L-n)}(E)}f(E).
\end{align}

\begin{figure}
    \centering
    \includegraphics[height=0.3\linewidth]{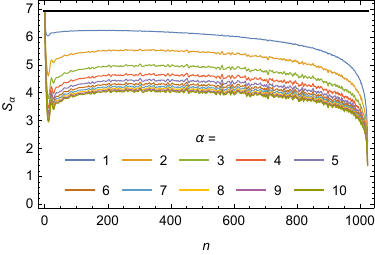}
    \caption{The $\alpha$-Renyi entropy $S_\alpha$ of the Krylov basis $\ke{\ket{O_n}}_{n=0}^{L-1}$ from the maximally entangled state in the GUE of $L=1024$ and $128$ realizations. The black line denotes $\log L$.}
    \label{fig:REc}
\end{figure}

Finally, a physical consequence of the confinement property discussed above is the decrease of entanglement in the Krylov basis from the maximally entangled state. Given a state $\ket{O_n}$ in the Krylov basis, we expand it on the equal-energy states as $\ket{O_n}=(1/\sqrt{L})\sum_{p=0}^{L-1} \psi_n(E_p)\ket{E_p,E_p}$, take the partial trace on $\H_R$, obtain a reduced density matrix in $\H_L$, and calculate the  $\alpha$-Renyi entropy $S_{\alpha}=\kd{\log( \sum_{p=0}^{L-1}\abs{\psi_n(E_p)}^{2\alpha}/L)}/(1-\alpha)$. We display the $\alpha$-Renyi entropies for all the states in the Krylov basis in the GUE in Fig.~\ref{fig:REc}. Recall that $\ket{O_0}$ is the maximally entangled state, whose $\alpha$-Renyi entropy is $S_\alpha=\log L,\,\forall \alpha$. When $n\gg1$, after the recurrence \eqref{recurrence} with random levels, the resulting Krylov state $\ket{O_n}$ has a nonuniform and random component on the equal-energy basis $\ket{E_p,E_p}$ with $E_p\in[-2\sqrt{1-n/L},2\sqrt{1-n/L}]$. This confinement of the spectrum leads to the decrease of entanglement for increasing $n$. In particular, when $1\ll n\ll L$, the dimension of the subspace spanned by the equal-energy basis within $[-2\sqrt{1-n/L},2\sqrt{1-n/L}]$ is exponentially large. So, first the entanglement decreases slowly. When $n$ approaches $L$, the dimension of the subspace shrinks quickly and the entanglement decreases quickly as well.

\subsection{Approximate the reduced spectral density and polynomials}\label{sec:ApproximatePolynomial}

We are constructing the reduced spectral density $\avg{\rho^{(L-n)}(E)}$ and the polynomials $\avg{\psi_n(E)}$ in the continuum limit.

\noindent{\bf Reduced spectral density}

\begin{figure}
    \centering
    \includegraphics[width=\linewidth]{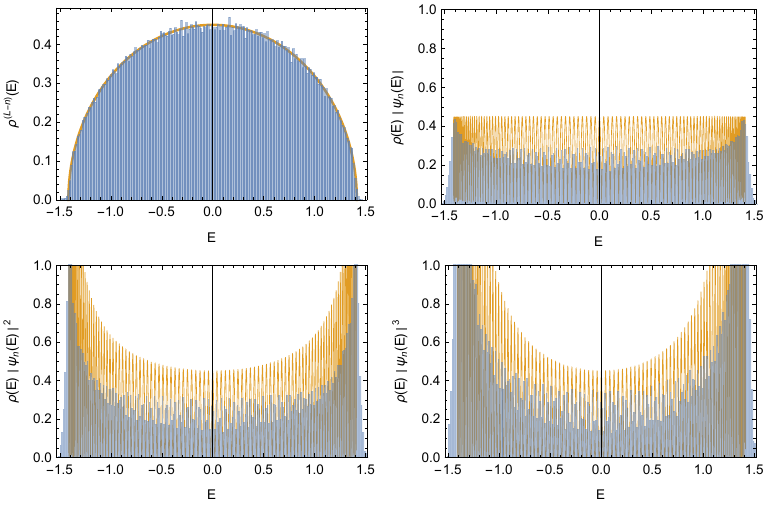}
    \caption{The reduced spectral density $\rho^{(L-n)}(E)$ (upper-left) and the spectral density $\rho(E)$ with weights $\abs{\psi_n(E)}^j$ (others), where $j=1,2,3,\, L=256,\, n=128$. The blue bins show the histograms of the average density functions over $1024$ realizations of the GUE. The orange curves show the reduced semicircle law \eqref{SubSemiCircle} and deformed Chebyshev polynomial \eqref{PolynomialChebyshev}. We use bin width $1/64$ in the histograms to show the oscillation.}
    \label{fig:PolynomialDensity}
\end{figure}

The reduced spectral density $\avg{\rho^{(L-n)}(E)}$ is determined by the spectrum of $\mathbf R^{(L-n)}$. We consider the limit value \eqref{LanczosLimit} for the right most Krylov chain of length $L-n$. From \eqref{LanczosToDensity}, it turns out to be a semicircle law as well
\begin{align}\label{SubSemiCircle}
    \rho_{\rm sc}^{(L-n)}(E)
    =\frac{L}{2\pi(1-n/L)}\sqrt{4\kc{1-\frac nL}-E^2}~\Theta\kc{4\kc{1-\frac nL}-E^2},
\end{align}
where we have introduced an additional normalization factor $1/(1-n/L)$ so that the reduced spectral density is normalized to $L$. 
This normalization will lead to the normalization \eqref{FixNorm} with a proper polynomial $\avg{\psi_n(E)}$. In the upper-left panel of Fig.~\ref{fig:PolynomialDensity}, we compare \eqref{SubSemiCircle} to the numerical result of the levels statistic in $\mathbf R^{(L-n)}$, which match very well.

\noindent{\bf Polynomials in the continuum limit}

We will construct the polynomials in the following two ways.
First, we calculate the polynomial $\avg{\psi_n(E)}$ in the continuum limit in the second-order formalism in Sec.~\eqref{sec:ContinuumLimitSecond}.
We consider $\epsilon=1/L$, finite $n/L$, and large $L$. By inserting the limit value \eqref{LanczosLimit} into the wave equation \eqref{WaveEq}, we find $V=O(1/L^4)$ which is negligible. 
With the boundary conditions \eqref{BoundaryCondition}, we obtain the ``polynomial'' in the continuum limit
\begin{align}\label{PolynomialContinuum}
    \psi_n^{\rm cont.}(E)=
    \begin{cases}
        (-1)^{n/2}\frac{\cos(EL(1-\sqrt{1-n/L}))}{(1-n/L)^{1/4}},
        & \text{even}\quad n\\
        (-1)^{(n-1)/2}\frac{\sin(EL(1-\sqrt{1-n/L}))}{(1-n/L)^{1/4}},
        &\text{odd}\quad n,
    \end{cases}
\end{align}
which coincides with the real part and imaginary part of \eqref{PolynomialContinuum1}. The frequency $L(1-\sqrt{1-n/L})$ in \eqref{PolynomialContinuum} is just the coordinate $y$ corresponding to $n$ in \eqref{WaveEq}. 
The transformation \eqref{PsiToPhi} from $E$ to $t$ on \eqref{PolynomialContinuum} gives a wave function $\phi_n(it)$. Since \eqref{PolynomialContinuum} is not a polynomial with finite degree and only contains two frequencies $\pm L(1-\sqrt{1-n/L})$, the wave function is localized around the characteristic curve $t=\pm L(1-\sqrt{1-n/L})$ in \eqref{CharacteristicCurves}, which is different from the numerical result in Fig.~\ref{fig:SPsGUEWave}, where the waves have long tails. Such difference in the tails will lead to a wrong slope behavior in the disconnected part of the transition probability. So we will only adopt it in the connected part of the transition probability.

Whatever, we find that $\psi_n^{\rm cont.}(E)$ is normalized to  
\begin{align}\label{ContinuumNorm}
\begin{split}
    &~\frac1{L}\int \psi_n^{\rm cont.}(E)^2\rho^{(L-n)}_{\rm sc}(E)\\
    =&~\frac1{2 \sqrt{1-n/L}}\kd{\, _0\tilde{F}_1\left(;2;-4 \left(n+L \left(\sqrt{1-n/L}-1\right)\right)^2\right)+1}    
\end{split}
\end{align}
with the reduced spectral density $\rho_{\rm sc}^{(L-n)}(E)$ in \eqref{SubSemiCircle}, where $\, _0\tilde{F}_1$ is the regularized confluent hypergeometric function. The normalization approaches $1/\kc{2 \sqrt{1-n/L}}$ when $n\gg1$. Consequently,  we will multiply \eqref{PolynomialContinuum} with $\sqrt2(1-n/L)^{1/4}$ when we use it to calculate the transition probability in \eqref{RampPlateauContinuum}.

\noindent{\bf Polynomials from the Chebyshev polynomials}

We can alternatively construct the polynomial $\psi_n(E)$ by deforming the Chebyshev polynomial of the second kind $U_n(E/2)$, which satisfies the orthogonality relation \eqref{Orthogonal} with respect to the measure given by the semi-circle law \eqref{SemiCircle}.
However, we show that $U_n(E/2)$ are the not the polynomials we need since they are given by constant Lanczos coefficients $a_n=0,\, b_n=1$ (see \eqref{ConstantPolynomial} in App.~\ref{sec:ConstantLanczos}). Here, the Lanczos coefficients are those given in \eqref{LanczosLimit}.
So, compared to the polynomials given by the Lanczos coefficients \eqref{LanczosLimit} with a plateau for $n\ll L$ and a descent for $n\lesssim L$, the Chebyshev polynomials $U_n(E/2)$ work for $n\ll L$ only.

We may modify the Chebyshev polynomials based on the above discussion of the matrix $\mathbf{\tilde L}$.
To this end, we approximate the sub-matrix $\mathbf L^{(n)}$ by replacing the Lanczos coefficients with their square mean value. 
For the limit value \eqref{LanczosLimit}, we set $a_m\to0$ and $b_m\to \bar b_n$, $\forall m \in [0,n-1]$ in $\mathbf L^{(n)}$. 
We may rescale the energy $E\to E/\bar b_n$ in the  Chebyshev polynomial $U_n(E/(2\bar b_n))$ such that the original domain of Chebyshev polynomials is rescaled to the domain $E\in[-2\bar b_n,2\bar b_n]$. 
Recall that the spectral domain of $\mathbf R^{(L-n)}$ is $[-2\sqrt{1-n/L},2\sqrt{1-n/L}]$. We will choose $\bar b_n=b_n=\sqrt{1-n/L}$ such that $U_n(E/(2\sqrt{1-n/L}))$ is approximately orthogonal to other $U_{n'}(E/(2\sqrt{1-n/L}))$ with $n'\sim n$ on the measure of $\rho_{\rm sc}^{(L-n)}(E)$.

However, we immediately encounter the following problem. $U_n(E/(2\sqrt{1-n/L}))$ around $E=0$ is proportional to $\exp(\pm iE(n+1)/(2\sqrt{1-n/L}))$. Again, by transforming $E$ to $t$, we find a shock wave propagating along $t=\pm (n+1)/(2\sqrt{1-n/L})$ that is very different from the characteristic curve $t=\pm L(1-\sqrt{1-n/L})$ found in the continuum limit for large $n$. 
To fix this problem, recalling that the spectral domain has been rescaled, we tune the degree of the Chebyshev polynomial to match the characteristic curve $t=\pm L(1-\sqrt{1-n/L})$. The resulting polynomial is 
\begin{align}\label{PolynomialChebyshev}
    &\avg{\psi_n(E)}\approx \psi_n^{\rm Cheb.}(E)=
    U_{d_n}\kc{\frac{E}{2 b_n}},\\
    &d_n=2L(1-\sqrt{1-n/L})\sqrt{1-n/L},\quad b_n=\sqrt{1-n/L},\nn
\end{align}
which will be confined in $E\in[-2\sqrt{1-n/L},2\sqrt{1-n/L}]$. The normalization coefficient $h_n$ is determined by \eqref{FixNorm}. The degree $d_n$ approaches $n$ for small $n$ and $0$ for $n=L$.
We compare the polynomials \eqref{PolynomialChebyshev} to the numerical result in Fig.~\ref{fig:Polynomials}.

By taking $f(E_p)=\delta(x-E_p)$ in \eqref{PolynomialEnsemble}, we get the distribution of $E_p$ with weight $\psi_n(E_p)^j$. 
We further compare the numerical statistics with the above weight $\abs{\psi_n(E_p)}^j$ to functions $\abs{\psi_n^{\rm Cheb.}(E)}^j\rho_{\rm sc}^{(L-n)}(E)$ for $j=1,2,3$ in Fig.~\ref{fig:PolynomialDensity}. 
We see that he approximate polynomials have closed moving average and oscillation frequency even at $n=L/2$ but their amplitudes and the edges behaviors are different. 
The numeric statistics give weaker oscillation mainly due to two reasons:
\begin{itemize}
    \item The approximation on $\mathbf L$ by $\tilde{\mathbf L}$ will affect the levels slightly such that some levels from $\mathbf L^{(n)}$ skip the zero points of $\psi_n(E_p)^2$. 
    \item The fluctuation of the polynomials $\psi_n(E_p)$ will also shift the phase of the oscillation. The mismatch between the phases in different realizations will suppress the oscillation in average.
\end{itemize}
The numeric statistic has smooth edges while the approximation has cliffy edges, where the discrepancy is due to our hard cut in breaking $\mathbf L$ to $\mathbf{\tilde L}$. The comparison in Fig.~\ref{fig:PolynomialDensity} shows that the normalization in \eqref{SubSemiCircle} is a good approximation for the spectral density with weight $\abs{\psi_n(E)}^j$.

\subsection{More on chaos in transition probabilities}\label{sec:ChaosBasis}

By considering the finite degree and the confinement of the polynomials, we improve the calculation of the transition probability for $n\lesssim L/2$ as compared to \eqref{RampFirstLook}.

We should first discuss the confinement effect in for the two-point function of the reduced spectral density. The selection in $\sum_p^{(L-n)}$ will not change the correlation between levels. We consider the following two-point function
\begin{align}\begin{split}\label{Sub2pt}
    &~\avg{\sum^{(L-n)}_{p,q}\delta(E'-E_p)\delta(E''-E_q)}\\
    =&~\avg{\rho^{(L-n)}(E')}\avg{\rho^{(L-n)}(E'')}+\avg{\rho^{(L-n)}(\bar E)}\kd{\delta(s)-\frac{\sin^2\kc{\pi\avg{\rho(\bar E)}s}}{\avg{\rho(\bar E)}(\pi s)^2}}
\end{split}
\end{align}
where $\bar E=(E'+E'')/2,\ s=E'-E''$.
We have introduced the improved sine kernel in \cite{Liu:2018hlr} to include the correlation between $E_p,\,E_q$ in the two delta functions. 
Since the sine kernel is localized in $-1/L<E_p-E_q<1/L$, we assume that $E_p$ and $E_q$ belongs to the summation $\sum_{p\neq q}^{(L-n)}$ or not simultaneously. So, only one reduced spectral density $\avg{\rho^{(L-n)}(\bar E)}$ is present with the sine kernel. If we integrate out $E''$, we reproduce the one point function \eqref{Sub1pt}. We can also use the box approximation by replacing $\avg{\rho(\bar E)}\to\avg{\rho(0)}$ in the sine kernel \cite{Liu:2018hlr}.

Similarly, by assuming the decorrelation between polynomials and spectral density after considering the confinement of $\psi_n(E_p)$, we approximate the transition probability as
\begin{subequations}
 \begin{align}
     &\avg{\abs{\phi_n(\beta+it)}^2}\nn\\
    =&\frac1{L^2}\int DE\sum_{p,q}\psi_n(E_p;\ke{E})\psi_n(E_q;\ke{E})e^{-(\beta+it) E_p-(\beta-it)E_q}\\   
    \approx&~\frac1{L^2}\int DE \sum_p^{(L-n)}\psi_n(E_p;\ke{E})^2 e^{-2\beta E_p}
    +\frac1{L^2}\int DE dE'dE''\\ &\hspace{0.05\linewidth} \psi_n(E';\ke{E})\psi_n(E'';\ke{E}) e^{-(\beta+it) E'-(\beta-it)E''}
    \sum^{(L-n)}_{p\neq q}\delta(E'-E_p)\delta(E''-E_q)\nn\\ 
    \approx&~\frac1{L^2}\int dE\avg{\psi_n(E)}^2 \avg{\rho^{(L-n)}(E)} e^{-2\beta E}
    +\abs{\frac{1}{L}\int dE' \avg{\psi_n(E')}\avg{\rho^{(L-n)}(E')} e^{-\tau E'}}^2 \nn\\
    &-\frac{1}{L^2}\int dE'dE''
    \avg{\psi_n(E')}\avg{\psi_n(E'')}
    \avg{\rho^{(L-n)}(\bar E)}
    \frac{\sin^2\kc{\pi\avg{\rho(\bar E)}s}}{\avg{\rho(\bar E)}(\pi s)^2}e^{-2\beta \bar E-its} \label{TransitionProbabilityApp}
 \end{align}
 \end{subequations}
where $\bar E=(E'+E'')/2$, and $s=E'-E''$. At the last step, we neglect the spectral correlation between the polynomials and delta functions. Although the sine kernel is localized in $-1/L<s<1/L$, we can not identify the $E'$ and $E''$ in the two polynomials, since their product, as a function of $s$, oscillates quickly when $n\lesssim L$. 

We have dropped the complicated correlation between the spectra in polynomials, namely
\begin{align}\label{ConnectedPolynomials}
    \frac1{L^2}\int DE dE'dE'' \psi_n(E';\ke{E})\psi_n(E'';\ke{E})e^{-(\beta+it)E'-(\beta-it)E''}\avg{\rho^{(L-n)}(E')}\avg{\rho^{(L-n)}(E'')},
\end{align}
which we are unable to calculated. However, we will see its contribution to the transition probability numerically. 

Finally, we will approximate the ensemble averages in \eqref{TransitionProbabilityApp} with the reduced spectral density $\rho_{\rm sc}^{(L-n)}(E)$ in \eqref{SubSemiCircle} and the deformed Chebyshev polynomial $\psi_n^{\rm Cheb.}(E)$  in \eqref{PolynomialChebyshev} or the polynomial in the continuum limit $\psi_n^{\rm cont.}(E)$ in \eqref{PolynomialContinuum}.
The transition probability is written as
\begin{align}
    &~\avg{\abs{\phi_n(\beta+it)}^2}\label{ProbabilitySplit}\\
    \approx&~
    \abs{\frac{1}{L}\int dE \avg{\psi_n(E)}\rho_{\rm sc}^{(L-n)}(E)e^{-(\beta+it)E}}^2+\frac1{L^2}\int dE\avg{\psi_n(E)}^2 \rho_{\rm sc}^{(L-n)}(E) e^{-2\beta E}  \nn\\
    &-\frac{1}{L^2}\int dEds \avg{\psi_n(E+s/2)}\avg{\psi_n(E-s/2)}\rho_{\rm sc}^{(L-n)}(E)
    \frac{\sin^2\kc{\pi\rho_{\rm sc}(E)s}}{\rho_{\rm sc}(E)(\pi s)^2}e^{-2\beta E-its}+\cdots \nn
 \end{align}

The first term of \eqref{ProbabilitySplit} contributes to the rise-slope. From our numerical results in Fig.~\ref{fig:SPsGUEWave}, we expect a sharp shock wave and a long tail on the Krylov chain, which corresponds to a sharp rise and a long slope along the real-time axis. The continuum limit fails to describe the long slope since it is localized at a frequency. We will adopt the deformed Chebyshev polynomial \eqref{PolynomialChebyshev} in the first term of \eqref{ProbabilitySplit}. The first term is the absolute square of
\begin{align}
    \avg{\phi_n(\tau)}
    \approx &~\frac{1}{2\pi\kc{1-n/L}}\int_{-2\sqrt{1-n/L}}^{2\sqrt{1-n/L}} dE~ U_{d_n}\kc{\frac{E}{2\sqrt{1-n/L}}} e^{-\tau E} \sqrt{4\kc{1-\frac nL}-E^2} \nn \\
    =&~ (-1)^n\frac{d_n+1}{\tau\sqrt{1-n/L}}I_{d_n+1}\kc{2\tau \sqrt{1-\frac nL}}  \label{rise-slope}
\end{align}
with $\tau=\beta+it$ and $I_n(x)$ the $n$-the modified Bessel function of the first kind.
So the transition probability $\avg{\abs{\phi_n(it)}^2}$ rises as $(1-n/L)^{d_n} t^{2 d_n}/\Gamma(d_n+1)^2$ initially, reaches its first peak around the time $L(1-\sqrt{1-n/L})$, and then decays as $(d_n+1)^2/\kd{\pi t^3(1-n/L)^2}$ over time.

The second term of \eqref{ProbabilitySplit} contributes to the plateau. From the normalization \eqref{FixNorm}, the plateau value is $1/L$.

\begin{figure}
    \centering
    \includegraphics[height=0.35\linewidth]{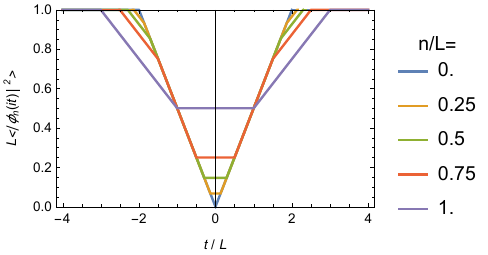}
    \caption{The ramp(-up)-plateau in the continuum limit \eqref{RampPlateauContinuum} as functions of $t/L$ for different $n/L$. We specially plot the negative axis of $t/L$ to show how the step in $-1+\sqrt{1-n/L}<t/L<1-\sqrt{1-n/L}$ emerges.}
    \label{fig:ramp-plateau}
\end{figure}

The third term of \eqref{ProbabilitySplit} contributes to the ramp-up behavior for $n\lesssim L/2$. The product $\avg{\psi_n(E+s/2)}\avg{\psi_n(E-s/2)}$ is an even and oscillating function of $s$ with a frequency of order $n$. Under the Fourier transformation of the level difference $s$, the oscillation will shift the time $t$ by some scales of $n$, which is not negligible when $n$ is comparable to $L$.
To estimate this effect analytically for $n\gg1$, we employ the polynomial in the continuum limit \eqref{PolynomialContinuum} and the reduced semicircle law \eqref{SubSemiCircle} with a normalization factor of $2\sqrt{1-n/L}$ from \eqref{ContinuumNorm}. To get a simple expression, we adopt the box approximation by sending $\rho_{\rm sc}(E)\to\rho_{\rm sc}(0)$ in the sine kernel.
The sum of the second and third terms for even $n$ at $\beta=0$ is given by
\begin{align}
    &~\frac1L-\frac{\pi}{4L(L-n)}
    \int_{-2\sqrt{1-n/L}}^{2\sqrt{1-n/L}} dE\int_{-\infty}^{\infty} ds~\sqrt{4\kc{1-\frac nL}-E^2}\frac{\sin^2\kc{Ls}}{(\pi s)^2}e^{-its}\label{RampPlateauContinuum}\\
    &~~~~
    \times \cos\kc{2L(E+s/2)(1-\sqrt{1-n/L})} \cos\kc{2L(E-s/2)(1-\sqrt{1-n/L})} \nn
    \\
    =&~\frac1{4L^2}\kd{\min\kc{\abs{t-L(1-\sqrt{1-n/L})},2L}
    +\min\kc{\abs{t+L(1-\sqrt{1-n/L})},2L}}.\nn
\end{align}
For odd $n$, we obtain the same result. The function is plotted in Fig.~\ref{fig:ramp-plateau}. The plateau value is always $1/L$, and the time dependence can be understood by the movement of the two $\min$ terms for different $n$. Each of them, as a function of $t$, is given by an upside-down triangle centered on the characteristic curve $t=\pm L(1-\sqrt{1-n/L})$, with a width of $4L$.
When $n=0$, the two triangles, centered at $t=0$, stack up to form a larger triangle, which corresponds to the ramp $t/(2L^2)$ of survival probability before the plateau time $t_p=2L$. When $n>0$, the two triangles move apart due to the oscillation of the polynomials on $s$. Then a step appears between the centers of the two triangles, whose height is $(1-\sqrt{1-x})/(2 L)$ and width is $2L(1-\sqrt{1-n/L})$. The ramp behaves as $t/(2L^2)$ when $L(1-\sqrt{1-n/L})<t<L(1+\sqrt{1-n/L})$, and as $t/(4L^2)$ when $L(1+\sqrt{1-n/L})<t<L(3-\sqrt{1-n/L})$. Thus, the plateau time is $t_p=L(3-\sqrt{1-n/L})$.

The dip time, which is the time when the ramp surpasses the slope, is given by $t_d\approx \sqrt[4]{2/\pi} \sqrt{L (n+1)}$ for $n\lesssim L/2$. Therefore, the ramp region of the transition probability $\avg{\abs{\phi_n}^2}$ lasts for $\sqrt{(L-n)L}$ long.

\begin{figure}
    \centering
    \includegraphics[height=0.4\linewidth]{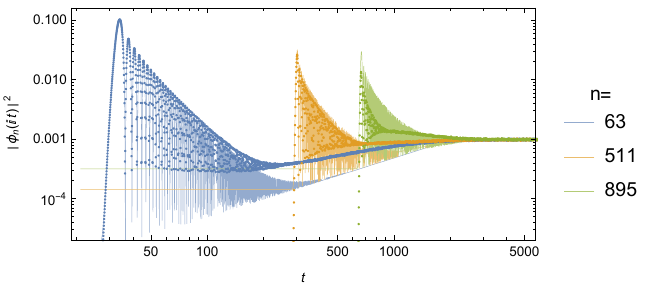}
    \caption{The transition probability $|\phi_n(it)|^2$ for $n=63,511,895$ as functions of time $t$, where $L=1024$. The dots are the numerical results in the average of $4096$ realizations and the curves are the analytical results from \eqref{rise-slope} and \eqref{RampPlateauContinuum}.}
    \label{fig:TPns}
\end{figure}

In Fig.~\ref{fig:TPns}, we compare the numerical results with the analytical results obtained from \eqref{rise-slope} and \eqref{RampPlateauContinuum}. The rise-slope behaviors match well until $n$ approaches $L$, but the ramp-up behaviors obtained from \eqref{RampPlateauContinuum} are smaller than the numerical results due to neglecting the fluctuations from the polynomials, as discussed in \eqref{ConnectedPolynomials} of App.~\ref{sec:PolynomialEnsemble}.

For $n\gtrsim L/2$, we are unable to reproduce the behavior of the transition probability analytically, including the rise-slope and ramp-down. This is because the Lanczos coefficients exhibit strong fluctuations at large $n\gtrsim L/2$, as shown in Fig.~\ref{fig:LanczosGUETFD}. Consequently, the polynomial $\psi_n(E_p;\ke{E})$ receives strong fluctuations from the spectrum $\ke{E}$, and the correlation between the two polynomials is not negligible. Such fluctuations in polynomials lead to strong fluctuating phenomena in the transition probability for $n\gtrsim L/2$ in both rise-slope behavior and ramp-down behavior. 

As an evidence of the strong fluctuation, in Fig.~\ref{fig:wave}, the transition probability in a single realization fluctuate strongly at large $n$. Also, in Fig.~\ref{fig:SPsGUETFD}, the connected part nearly dominates the whole transition probability in the ensemble average, which implies strong fluctuations in each realization.

A possible explanation of the ramp-down behavior for $n\gtrsim L/2$ is as follows. The ramp-down behavior could be originated from the slope but with strong fluctuation from the polynomials. The fluctuation in the polynomials leads to phase fluctuation in the oscillation of the slope. For example, we can imagine a phase $\varphi$ fluctuation in $\abs{I_{d_n+1}(2it\sqrt{1-n/L}+\varphi)}^2$ of \eqref{rise-slope}. In the ensemble average, the oscillations in the slope between different realization are cancelled with each other due to the phase fluctuation and end up with a moving average. The moving average of the slope behaves like the ramp down behavior in the numerical average.  

\section{Constant Lanczos coefficients}\label{sec:ConstantLanczos}

Here we give a simple example for the application of our Ehrenfest theorem and show several typical features of Krylov complexity. We consider the case of constant Lanczos coefficients in the interior
\begin{align}\label{ConstantLanczos}
	a_n=0,\quad b_n=1,
\end{align}
and $b_0=b_L=0$ at the chain endpoints. Without loss of generality, we work with dimensionless quantities, so that the spectrum $\ke{E_p}$, the Lanczos coefficients $\ke{a_n,b_n}$, and time $\tau=\beta+it$ are dimensionless.

The $b_{n+1}^2-b_n^2$ and $a_{n+1}-a_n$ terms in the Ehrenfest theorem \eqref{Ehrenfest} vanish except $b_1^2-b_0^2=1$ and $b_L^2-b_{L-1}^2=-1$. So the growth of Krylov complexity is driven by the probability at the two endpoints, namely,
\begin{align}\label{EhrenfestConstantLanczos}
    \partial_t^2 K(t;\beta)
    =2\frac{\abs{\phi_0(\beta+it)}^2-\abs{\phi_{L-1}(\beta+it)}^2}{S(2\beta)}.
\end{align}

Eq.~\eqref{ConstantLanczos} effectively describes the flat Lanczos coefficients in a chaotic system at the $n\ll L$ limit. Since the wave function is initially localized at $n=0$, the short time evolution of the wave function as well as the Krylov complexity in a general system is effectively described by the first term of Eq.~\eqref{EhrenfestConstantLanczos}.

\subsection{Eigenstate}

From \eqref{polynomial}, we find that the orthogonal and normalized polynomial is given by the Chebyshev polynomial of second kind \cite{Dymarsky:2019quantum,Muck:2022xfc}
\begin{align}\label{ConstantPolynomial}
	\psi_n(E)=U_n(E/2).
\end{align}
The eigenenergy and the measure are
\begin{align}
	E_p=2\cos\frac{\pi (L-p)}{L+1},\quad 
    \abs{\avg{E_p|0}}^2
    =\frac{4-E_p^2}{2(L+1)},\quad 
    p=0,1,2,\cdots,L-1.
\end{align}	

At low temperature, according to \eqref{Complexitylow}, the complexity will converge to
\begin{align}
	K_\text{low}(\beta)\approx 
	\frac{L-1}{2},
\end{align}
where the tedious sub-leading term is not shown. It describes the $\beta\gg L$ limit.

At the long-time average, according to \eqref{ComplexityAverage}, the plateau value of the complexity is given by
\begin{align}
	K_\infty(\beta)=\frac {L-1}2.
\end{align}

\subsection{Large dimension limit}

\begin{figure}
	\centering
	\includegraphics[width=0.49\linewidth]{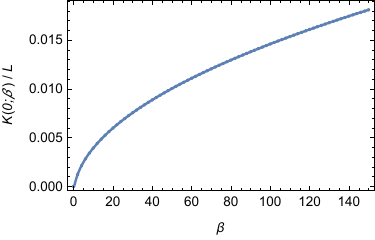}
	\includegraphics[width=0.49\linewidth]{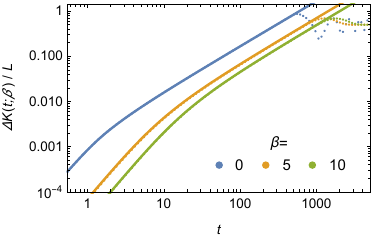}
	\caption{The Krylov complexity as a function of inverse temperature (left) and time (right) in the case of constant Lanczos coefficient with $L=1024$. The dots represent the numerical results and the solid curves represent analytical approximation in \eqref{KCConstantBeta} and \eqref{KCConstant}.}
	\label{fig:KCConstant}
\end{figure}

In the large $L$ limit, we can calculate the evolution of wave function and complexity analytically. The density of state and the measure are approximately
\begin{align}\label{ConstantDensity}
	\rho(E)\approx\frac{L}{\pi\sqrt{4-E^2}},\quad \abs{\avg{E|0}}^2\approx\frac{4-E^2}{2L}.
\end{align}
The wave function is given by the integral
\begin{align}\label{WaveFunctionConstant}
	\phi_n(\tau)\approx
	\frac{1}{2\pi}\int_{-2}^2  \psi_n(E)e^{-\tau E} \sqrt{4-E^2} dE
	=(-1)^n\frac{n+1}{\tau}I_{n+1}(2\tau ),
\end{align}
where $I_n$ is the $n$-th modified Bessel function of the first kind. $\phi_n(\tau)$ becomes nonzero for all $n$ in this large $L$ limit. We can calculate the survival amplitude and Krylov complexity along imaginary time 
\begin{align}
	S(2\beta)&
	\approx \frac{I_1(4 \beta )}{2\beta },\\
	K(0;\beta)&
	\approx \frac{2\beta  \left(I_0(2\beta ){}^2+I_1(2\beta ){}^2\right)}{I_1(4\beta )}-1 
	\approx 2\sqrt\frac{2\beta }{\pi }-1 ,\ \text{when}\ \beta\gg1, \label{KCConstantBeta}
\end{align}
where we have used the recursion relations
$2n I_n(x)=x(I_{n-1}(x)-I_{n+1}(x))$ and $2I_n'(x)=I_{n-1}(x)+I_{n+1}(x)$.
Along the real time, we use \eqref{EhrenfestConstantLanczos} and find
\begin{align}\label{ddKChebyshev}
	\partial_t^2 K(t;\beta)
	\approx2\frac{2\beta}{\beta^2+t^2}\frac{\abs{I_1(2\beta+i2t)}^2}{I_1(4\beta)}.
\end{align}
We integrate over the real time and obtain the complexity different $\Delta K(t;\beta)$ 
\begin{align}\label{KCConstant}
    \Delta K(t;\beta)\approx
    \begin{cases}
        _1F_2\left(-1/2;1,2;-4t^2\right)-1\approx \frac{16}{3 \pi }t, & \beta=0 \\
        \sqrt{\frac{8}{\pi\beta}} \left(\sqrt{\beta ^2+t^2}-\beta \right)
        \approx 
        \sqrt{\frac{8}{\pi\beta}} t, & \beta\gg1
    \end{cases},
\end{align}
where $F$ is the generalized hypergeometric function and we show the linear growth at the last step.
The above results show the typical square-to-linear-to-plateau growth of Krylov complexity along the real time at finite temperature. We compare these formulas to the numeric results in Fig.~\ref{fig:KCConstant}.

\subsection{Finite dimension effects}

We would like to study the behavior of the Krylov complexity near the plateau time. We will consider a large but finite $L$, such that $\beta,t>L$ is accessible. For simplicity, we will adopt the continuous Schr\"odinger equation \eqref{cSchEq} with potential $\tilde V(y)=-2 \Theta(y)\Theta(L-y)$ and the boundary conditions $\Phi(0)=\Phi(L)=0$. The eigensystem are
\begin{align}
    E_k=\kc{\frac{k\pi}{L}}^2-2,\quad 
    \Phi_k(y)=\sqrt{\frac{2}{L}} \sin \frac{\pi k y}{L},\quad
    k=1,2,...,\tilde L
\end{align}
where $k$ is originally bounded by the dimension of the Krylov space $L$, but we have released this constraint by consider the bound $\tilde L$ instead and will sent $\tilde L\to\infty$ finally for simplicity. 
We consider the initial state $\tilde\varphi(\tau=0)=\delta(y-\epsilon)$ locating beside the boundary with small distance $\epsilon$. Its overlap to the eigenstate is $\avg{E_k|\tilde\varphi(0)}=\sqrt{2} \pi k\epsilon/L^{3/2}$. 
So the state evolves as
\begin{align}
    \tilde\varphi(\tau)=\frac{2 \pi  \epsilon^2}{L^2}\sum_{k=1}^{\tilde L} k e^{-E_k\tau} \sin \frac{\pi k y}{L}, 
\end{align}
Since the Lanczos coefficients are flat, the Krylov complexity $K$ is the expectation value of the position operator $y$. The complexity operator on the energy basis are
\begin{align}
    K_{kl}=\avg{E_k|y|E_l}=
	\begin{cases}
		\frac L2,	&	k=l,\\
		-\frac{8 L kl}{\pi ^2 \left(k^2-l^2\right)^2}, & \text{odd}\ k+l \\
		0,	& \text{others}
	\end{cases}.
\end{align}
Its expectation value is
\begin{align}
    \avg{K}
    =&~\frac{\sum_{kl} \avg{\tilde\varphi(\tau)|E_k} K_{kl} \avg{E_l|\tilde\varphi(\tau)}}{\avg{\tilde\varphi(\tau)|\tilde\varphi(\tau)}} \nn \\
    =&~\frac L2-\frac{\epsilon^2}{L^2S(2\beta)}\\
    &~\times\sum_{kl;\, \text{odd}~k_+}
    \kc{\frac{k_-}{k_+}-\frac{k_+}{k_-}}^2
    \exp \left\{-\frac{\pi^2}{2 L^2}\left[\beta  \left(k_-^2+k_+^2-\frac{8L^2}{\pi^2}\right)+2 i k_- k_+ t\right] \right\},\nn
\end{align}
where $\tau=\beta+it,\ k_\pm=k\pm l$, and
\begin{align}
    S(2\beta)=&~\avg{\tilde\varphi(\beta)|\tilde\varphi(\beta)}
    =\frac{2 \pi ^2  \epsilon ^2}{L^3}\sum_k k^2e^{\beta  \left(4-\frac{2 \pi ^2 k^2}{L^2}\right)} \nn \\
    \approx&~ \frac{e^{4 \beta } \epsilon ^2}{8 \beta ^{3/2}}\left(\sqrt{\frac{2}{\pi }} \text{erf}\left(\pi  \sqrt{2\beta }\right)-4 e^{-2 \pi ^2 \beta } \sqrt{\beta }\right).
\end{align}
By taking the summation, one can get the value of Krylov complexity for general $\beta$ and $t$.

\bibliographystyle{JHEP}
\bibliography{refs}

\end{document}